\documentclass[12pt]{iopart}

\newcommand{\be}{\begin{eqnarray}}
\newcommand{\ee}{\end{eqnarray}}
\newcommand{\bea}{\begin{eqnarray}}
\newcommand{\eea}{\end{eqnarray}}

\usepackage{graphicx}
\usepackage{bm}
\usepackage{color}
\usepackage{slashed}
\usepackage{cite}
\usepackage{graphicx}
\usepackage{multicol}
\usepackage{graphicx}
\usepackage{color}
\usepackage{slashed}
\usepackage{CJKutf8}
\begin{document}
\begin{CJK}{UTF8}{<font>}
\title{Influence of accretion disk on the optical appearance of the Kazakov-Solodukhin black hole}

\author{Yu-Xiang Huang$^{1}$, \ Sen Guo$^{*2}$, \ Yu-Hao Cui$^{3}$,  \ Qing-Quan Jiang$^{*1}$, \ Kai Lin$^{*3}$}

\address{
$^1$School of Physics and Astronomy, China West Normal University, Nanchong 637000, People's Republic of China\\
$^2$Guangxi Key Laboratory for Relativistic Astrophysics, School of Physical Science and Technology, Guangxi University, Nanning 530004, People's Republic of China\\
$^3$Hubei Subsurface Multi-scale Imaging Key Laboratory, School of Geophysics and Geomatics, China University of Geosciences, Wuhan 430074, People's Republic of China}

\ead{yxhuangphys@126.com}
\vspace{10pt}
\begin{indented}
\item[]Oct. 2023
\end{indented}

\begin{abstract}
The optical characteristics of the Kazakov-Solodukhin (KS) black hole (BH) surrounded by a thin disk is investigated. By utilizing the ray-tracing technique, we derive the the direct and secondary images to illustrate the influence of observational inclination angle ($\theta_{0}$) and quantum correction parameter ($a$) on the observable properties. Our calculation involves determining the effective potential, light deflection, and azimuthal angle of the BH, as well as the radiation flux of the accretion disk. By assuming three function models, we determine that the exponential function is the most suitable for describing the relationship between the observed flux and the quantum deformation parameter. Our research reveals that the observable characteristics of a BH are affected not only by the shape of the accretion disk, but also by its spatiotemporal properties and the observer's inclination angle. We observe that a BH encircled by an optically thick accretion disk yields different results from those of an optically thin disk.
\end{abstract}

\section{Introduction}
\label{intro}
\par
Black hole (BH), which is regarded as the most mysterious compact object in the Universe, have received a large amount of observational evidence to date. The Event Horizon Telescope (EHT) has been instrumental in capturing the images of the supermassive BH at the center of the Messier 87$^{*}$ (M87$^{*}$) galaxy \cite{1}. Additionally, the EHT published findings on the magnetic field structure of the surrounding accretion disk in 2019, illustrating that the magnetically arrested accretion disk is consistent with the predictions of general relativistic magnetohydrodynamic (GRMHD) models \cite{2}. Subsequently, the EHT captured the inaugural horizon-scale radio observations image of the Sagittarius A$^{*}$ (Sgr A$^{*}$) positioned at the heart of the Milky Way, and various imaging and modeling analyses have supported the image being primarily composed of a bright, thick ring with a diameter of $51.8 {\pm} 2.3$ ${\rm {\mu as}}$ \cite{3}. Given these discoveries, it is reasonable to hypothesize that the accretion disk surrounding the BH could be influencing the observational characteristics of the BH shadow.

\par
If the accretion rate of a BH surpasses that of other celestial bodies in the galaxy, it can lead to the formation of an active galactic nucleus (AGN) through standard, radiation-efficient, thin disk accretion at the Eddington limit. Research on BH accretion disks dates back to the 1970s, the Shakura-Sunyaev model of the accretion disk describes it as geometrically thin and optically thick, providing a description of its geometric thickness and optical properties \cite{4}. By considering the thin accretion disk model, Luminet obtained the direct and secondary images of the accretion disk through semi-analytic method, demonstrating that the simulated photograph is consistent with BHs of any mass accreting matter at any moderate rate \cite{5}. Notably, this method uses elliptic integrals for analysis. Utilizing the Luminet's method, Gyulchev $et~al.$ analyzed the lensing properties of the Janis-Newman solution, as well as the stable circular geodesics in the equatorial plane. They obtained the optical appearance and the apparent radiation flux of a thin accretion disk around the static Janis-Newman-Winicour naked singularity \cite{6}. Furthermore, Paul $et~al.$ employed two distinct numerical methods to generate images of thin accretion disks situated in the background of Kerr-like and Teo class wormholes \cite{7}. These results were extended to a rotating traversable wormhole of Teo class and the Schwarzschild BH pierced by a cosmic string \cite{8,9}. BH accretion disk can account for the spectral characteristics of many AGNs, particularly the redshift of the $Fe$ $K\alpha$ line. In Tanaka $et~al.$'s research, it was shown that the iron emission is influenced by gravitational redshift \cite{10}. Cunningham used the ray-tracing techniques to study the accretion disk around the Kerr black hole, and found that different observation angles have a greater impact on the observed radiation redshift \cite{11}. Based on the consideration of the thin accretion disk surrounding the Schwarzschild BH, the image of the disk can be accurately characterized using ray-tracing numerical methods and radiation transfer \cite{12}. Extensive research on the image of the disk in various BH modified gravity contexts has also been discussed in literature \cite{13,14,15,16,17,18,19,20,21,22,23}.

\par
The investigation of singularities constitutes a vital area of focus in the field of general relativity (GR). It is not only one of the earliest applications of the energy condition, but also functions as a benchmark for assessing the completeness of GR. The existence of singularities results in the incompleteness of geodesics, which in turn highlights the need for quantum gravity to address this pressing issue \cite{24}. Investigations have been conducted on the quantum effects in the vicinity of singular points, event horizons, and spherically symmetric gravitational collapses \cite{25,26}. The findings indicate that quantum effects may potentially prevent the emergence of intrinsic singularities at the origin. Extensive studies have been conducted on the quantum effects on BH spacetime in various contexts \cite{27,28,29,30,31}. Hooft investigated quantum gravity within the semi-classical method framework and discovered that BH production accompanied by the coherent emission of real gravitons occurs at energies significantly higher than the Planck mass \cite{32}. On the other hand, the BH solution in the context of quantum gravity remains an open question. The deformation situation of the Schwarzschild BH solution caused by matter field were investigated by Kazakov $et~al.$, who argue that this quantum correction solution could be describe the quantum fluctuations in the field of matter \cite{33}. Based on this groundbreaking exploration, Konoplya utilized the computation of quasinormal modes to further investigate this BH solution and discovered that turning on quantum deformation results in a decrease in the radius of the BH's shadow \cite{34}. Xu $et~al.$ investigated the weak and strong deflection gravitational lensing by Kazakov-Solodukhin (KS) BH and reported observable effects \cite{35}. Additionally, KS BH has been used in many other areas of research \cite{36,37,38,39}.

\par
It remains currently unclear whether quantum corrections, an important aspect of theoretical BH solutions, can impact the observable properties of BHs. To investigate this issue, Peng $et~al.$ examined the shadow of a quantum-corrected Schwarzschild BH in the context of an optically and geometrically thin disk. Their study demonstrated that due to quantum corrections, the photon sphere and critical curve violate the general inequality of asymptotically flat BHs adhering to the zero energy condition in Einstein's gravity. These violations result in observable changes in the appearance of the accretion disk around the BH \cite{40}. In this analysis, we extend the study by examining the optical appearance of a KS BH with a geometrically thin and optically thick accretion disk. Note that our's work is an extension of \cite{40}, with the goal of providing a more comprehensive understanding of the possible optical appearance of the target BH. Specifically, we analyze the direct and secondary images of the accretion disk and investigate the impact of the deformation parameter on the radiation flux. Additionally, we consider the distribution of redshift and flux in the images.

\par
The paper is structured as follows. Section \ref{sec:2} discusses the effective potential of the KS BH and limits its deformation parameter using EHT observations. In Section \ref{sec:3}, we analyze the orbits of the direct and secondary images of the KS BH and investigate the unit flux distribution of radiation energy on a disk, as well as the functional relationship between its flux and the deformation parameter. Section \ref{sec:4} considers the red shift into its radiant energy flux and presents its empirical function. Finally, we conclude in Section \ref{sec:5}.

\section{The effective potential of KS BH}
\label{sec:2}
\par
The KS BH metric can be written as \cite{33}
\begin{equation}
\label{2-1}
{\rm d}s^{2}=-f(r){\rm d}t^{2}+\frac{1}{f(r)}{\rm d}r^{2}+r^{2}{\rm d}\theta^{2}+r^{2}\sin^{2}\theta {\rm d}\phi^{2},
\end{equation}
where $f(r)$ is the metric potential,
\begin{equation}
\label{2-2}
f(r)=\frac{\sqrt{r^{2}-a^{2}}}{r}-\frac{2M}{r},
\end{equation}
in which $M$ represents the BH mass. The parameter $a$ arises from the effective two-dimensional dilaton gravity and serves as a deformation parameter. When $a$ is equal to zero, Eq. (\ref{2-2}) simplifies the Schwarzschild BH. The horizon radius of this BH can be expressed as
\begin{equation}
\label{2-3}
r_{\rm H}=\sqrt{a^{2}+4M^{2}}.
\end{equation}
Note that the curvature of the two-dimensional sphere $r=a$ diverges in a four-dimensional image, and the extremal value problem presented in the above equation can be resolved \cite{33}.

\par
In order to study the optical appearance of the KS BH, we must know how photons move around it. The motion of photons can be represented by the Euler-Lagrange equation
\begin{equation}
\label{2-4}
\frac{{\rm d}}{{\rm d}\lambda}\Big(\frac{\partial {\rm \mathcal{L}}}{\partial \dot{x}^{\rm \mu}}\Big) = \frac{\partial {\rm \mathcal{L}}}{\partial x^{\rm \mu}},
\end{equation}
where $\lambda$ is the affine parameter and $\dot{x}^{\mu}$ is the four-velocity of the photon. If we only consider the motion of photons on the equatorial plane, the Lagrange equation can be re-expressed as
\begin{equation}
\label{2-5}
\mathcal{L}=-\frac{1}{2}g_{\rm \mu \nu}\frac{{\rm d} x^{\rm \mu}}{{\rm d} \lambda}\frac{{\rm d} x^{\rm \nu}}{{\rm d} \lambda}=0.
\end{equation}
The generalized momentum $p_{\rm \mu}$ of a particle can be represented as follows:
\begin{equation}
\label{2-6}
 p_{\rm \mu}=\frac{\partial {\rm \mathcal{L}}}{\partial {\dot{x}^{\rm \mu}}}=g_{\rm \mu \nu}\dot{x}^{\rm \nu},
\end{equation}
which leads to the derivation of four equations of motion for a particle having energy $E$ and angular momentum $L$, i.e.,
\begin{eqnarray}
\label{2-7}
&&p_{\rm t}=g_{\rm t t}\dot{t}=-E,\\
\label{2-8}
&&p_{\rm \phi}=g_{\rm \phi \phi}\dot{\phi}=-L,\\
\label{2-9}
&&p_{\rm r}=g_{\rm r r}\dot{r},\\
\label{2-10}
&&p_{\rm \theta}=g_{\rm \theta \theta}\dot{\theta}.
\end{eqnarray}
Therefore, the motion equation of the null geodesic can be written as
\begin{equation}
\label{2-11}
\frac{p_{\rm t}^{2}}{f(r)}-\frac{p_{\rm \phi}^{2}}{r^{2}}=\frac{(p^{\rm r})^{2}}{f(r)}.
\end{equation}
Using the Eq. (\ref{2-11}), the radial component of the null geodesic is derived, i.e.
\begin{equation}
\label{2-12}
p^{\rm r}={\pm} E \sqrt{1-\frac{b^{2}}{r^{2}}f(r)},
\end{equation}
where $b \equiv L/E$ is the impact parameter, the symbol ``$\pm$'' indicates the direction of photon motion, either clockwise ($+$) or counterclockwise ($-$). The effective potential of the KS BH is obtained, we have
\begin{equation}
\label{2-13}
{\rm V}_{\rm eff}=r^{4}\Bigg(\frac{1}{b^{2}} - \frac{f(r)}{r^{2}}\Bigg).
\end{equation}
If the effective potential satisfies ${\rm V}_{\rm eff}=\frac{1}{b_{\rm c}^{2}}$ and ${\rm V}_{\rm eff}'=0$, the stability region of the photon orbit, namely the photon ring, can be determined based on the effective potential. For the KS BH, the photon ring radius and critical impact parameter can be expressed as
\begin{eqnarray}
\label{2-14}
&&r_{\rm ph}=\sqrt{\frac{3(\sqrt{9+2{x}^{2}}+3+{x}^{2})}{2}}M,\\
\label{2-15}
&&b_{\rm c}=\sqrt{\frac{\sqrt{27(\sqrt{9+2{x}^{2}}+3+{x}^{2})^{3}}}{2\sqrt{9+{x}^{2}+3\sqrt{9+2{x}^{2}}}-4\sqrt{2}}}M,
\end{eqnarray}
in which ${x}=\frac{a}{M}$ is a dimensionless parameter.

\par
Note that the KS BH shadow radius is affected by the quantum deformation parameter $a$. In terms of observation, the BH shadow diameter $d_{\rm sh}$ can be measured with the EHT data. As we all known that the angular size of the shadow of M87$^{*}$ is $\delta=(42 {\pm} 3)$ ${\rm {\mu as}}$, its distance is $D=16.8_{-0.7}^{+0.8}$ ${\rm Mpc}$, and BH mass is $M=(6.5 {\pm} 0.9) \times 10^{9}M_{\odot}$. Using this observation data, the diameter of its shadow can be calculated $d_{M87^{*}}=\frac{D\delta}M\simeq 11.0\pm 1.5$ \cite{41,42}.

\par
For Sgr A$^{*}$ BH, the EHT not only measured the angular size of the emission ring as $\delta_{\rm d}=(51.8 {\pm} 2.3)$ ${\rm {\mu as}}$ but also estimated the shadow angular $\delta=(48.7 {\pm} 7)$ ${\rm {\mu as}}$. The distance and BH mass of Sgr A$^{*}$ are also given separately as $D=(8.15 {\pm} 0.15)$ ${\rm kpc}$ and $M=(4.0_{-0.6}^{+1.1}) \times 10^{6}M_{\odot}$ \cite{3}. Therefore, the diameter of its shadow $d_{\rm SgrA^{*}}$ could be got.
Based on the aforementioned observation data, we constrain $a$ with EHT observations. As illustrated in Fig. 1, the result of KS BH is consistent with that derived from the EHT observations within the observational uncertainty. The results indicate that using the $1\sigma$ and $2\sigma$ confidence intervals of the $d_{\rm M87^{*}}$, the deformation parameter can be constrained as $a \leq 2.51$ within $1\sigma$ and $a \leq 3.39$ within $2\sigma$. When considering the confidence intervals of $d_{\rm sgrA^{*}}$, it can be constrained as $a \leq 0.55$ within $1\sigma$ and $a \leq 1.27$ within $2\sigma$.

\begin{center}
\includegraphics[width=7.5cm,height=5cm]{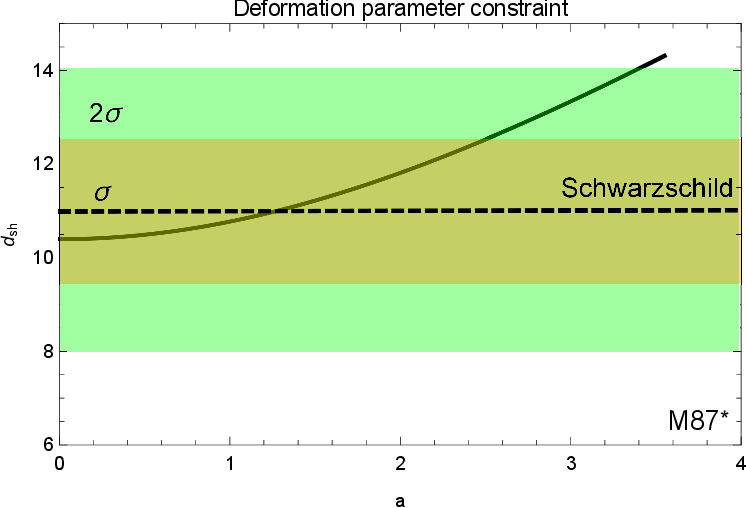}
\includegraphics[width=7.5cm,height=5cm]{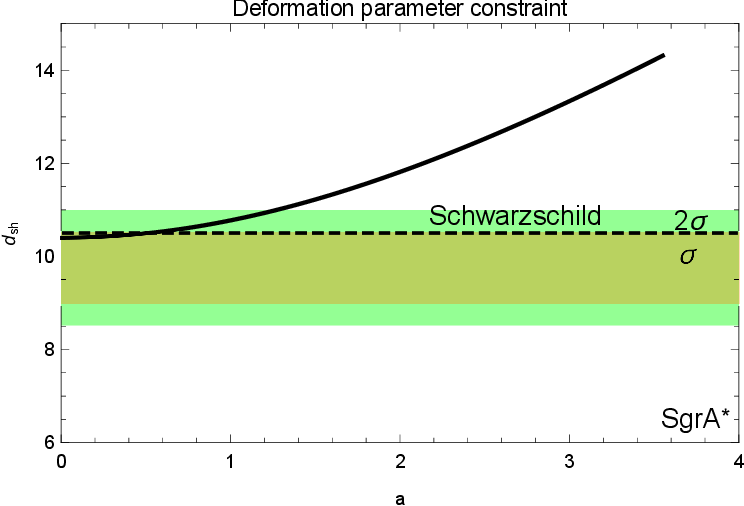}
\parbox[c]{15.0cm}{\footnotesize{\bf Fig~1.}  
Shadow diameter of the KS BH as a function of  the deformation parameter. The green and brown shaded regions represent the regions of $1\sigma$ and $2\sigma$ confidence intervals, respectively, with respect to the M87$^{*}$ and SgrA$^{*}$ observations.}
\label{fig1}
\end{center}

\par
Then, we derive the deflection angle of light moving around the KS BH. Based on Eqs. (\ref{2-7}) - (\ref{2-11}), we have
\begin{equation}
\label{2-16}
\Omega(u) \equiv \Big(\frac{{\rm d} u}{{\rm d} \phi}\Big)^{2} = 2 M u^{3} + \frac{1}{b^{2}}- u^{2}\sqrt{1-a^{2}u^{2}},
\end{equation}
where the $u$ is defined as $u \equiv 1/r$. If the impact parameter satisfies $b > b_{\rm c}$, the light rays will be deflected by the gravity effect, but will not be absorbed into the BH. When the impact parameter satisfies $b < b_{\rm c}$, light rays are drawn into BH in this situation. By using the Cardano formula, the Eq. (\ref{2-16}) can be re-written as follows:
\begin{eqnarray}
\label{2-17}
&&2 M u^{3} + \frac{1}{b^{2}}- u^{2}\sqrt{1-a^{2}u^{2}} \equiv 2 M G(u)\nonumber\\
&&= 2M (u-u_{1})(u-u_{2})(u-u_{3}).
\end{eqnarray}
Here, the cubic polynomial $G(u)$ has two positive roots and one negative root, satisfying $u_{1} \leq 0 < u_{2} < u_{3}$. Following the Ref. \cite{5}, these three roots can be represented by introducing a parameter, that is $Q^{2} \equiv (P-2M)(P+6M)$. Utilizing the periastron distance $P$ (a given periastron distance $P$ can be used to determine the corresponding impact parameter $b$ at infinity), we have
\begin{equation}
\label{2-18}
u_{1}=\frac{P-2M-Q}{4MP},~~u_{2}=\frac{1}{P},~~u_{3}=\frac{P-2M+Q}{4MP},
\end{equation}
and
\begin{equation}
\label{2-19}
b^{2}=\frac{1}{\frac{P-2M}{P^{3}}+u^{2}(\sqrt{1-a^{2}u^{2}}-1)}.
\end{equation}
Hence, the bending angle of light can be expressed as
\begin{equation}
\label{2-20}
\psi (u) = \sqrt{\frac{2}{M}} \int_{0}^{u_{2}} \frac{{\rm d}u}{\sqrt{(u-u_{1})(u-u_{2})(u-u_{3})}} - \pi.
\end{equation}
Obviously, it is difficult to directly integrate the above equation, so we refer to the semi analytical method of Luminet and convert this equation into the elliptic integral, we can obtain the following equation:
\begin{equation}
\label{2-21}
\psi (u) = \sqrt{\frac{2}{M}} \Bigg(\frac{2 F(\Psi_{1},k)}{\sqrt{u_{3}-u_{1}}} - \frac{2 F(\Psi_{2},k)}{\sqrt{u_{3}-u_{1}}}\Bigg) - \pi,
\end{equation}
where $\Psi_{1}=\frac{\pi}{2}$, $\Psi_{2}=\sin^{-1}\sqrt{\frac{-u_{1}}{u_{2}-u_{1}}}$, and $k=\sqrt{\frac{u_{2}-u_{1}}{u_{3}-u_{1}}}$. With the help of Eq. (\ref{2-18}), the total change of bending angle of the KS BH is calculate, i.e.,
\begin{equation}
\label{2-22}
\psi (u) = 2 \sqrt{\frac{P}{Q}} \Big(K(k)- F(\Psi_{2},k)\Big) - \pi,
\end{equation}
in which $K(k)$ is the complete elliptic integrals of the first kind.

\section{Trajectory and radiation intensity of the KS BH}
\label{sec:3}
\par
In this section, we consider the KS BH, which is surrounded by an optically thick and geometrically thin accretion disk. We investigate the direct and secondary images of the KS BH using the Luminet method. Furthermore, to clearly describe this model, we provide a simple observation diagram (see Fig. 2). For a given emitter $M$ with the coordinates $(r,\varphi)$, the observer can observe it at point $m$ on the observation plane $(b,\alpha)$. By changing the position of the emitter $M$, we can obtain the direct image with the coordinates $(b^{(d)},\alpha)$ and the secondary image with coordinates $(b^{(s)},\alpha+\pi)$. (These are described in detail in Ref. \cite{5}).
\begin{center}
\includegraphics[width=6.5cm,height=5cm]{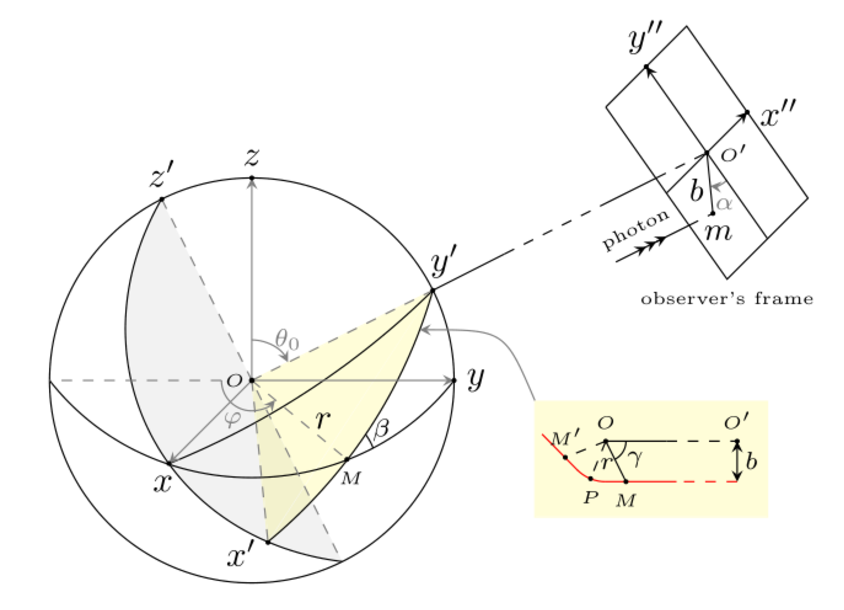}
\parbox[c]{15.0cm}{\footnotesize{\bf Fig~2.}  
The coordinate system is indicated by Ref. \cite{5}.}
\label{fig2}
\end{center}

\subsection{Ray trajectory of the KS BH}
\label{sec:3-1}
The angle of deflection required to reach the observer from point $M$ is $\gamma$, and the observer's inclination angle is denoted as $\theta_{0}$. By using the spherical triangle sine theorem, we have
\begin{equation}
\label{3-1}
\cos \alpha = \cos \gamma \sqrt{\cos^{2} \alpha + \cot^{2} \theta_{0}}.
\end{equation}
We can use elliptic integral to rewrite Eq. (\ref{2-17}) to get the direct image of accretion disk around the KS BH
\begin{eqnarray}
\label{3-2}
\gamma &&= \frac{1}{\sqrt{2M}} \int_{0}^{1/r}\frac{1}{\sqrt{G(u)}}{\rm d}u \nonumber\\
       &&= 2 \sqrt{\frac{P}{Q}} \Big(F(\zeta_{\rm r},k) - F(\zeta_{\rm \infty}, k)\Big),
\end{eqnarray}
where $F(\zeta_{\rm r},k)$ and $F(\zeta_{\rm \infty})$ are the elliptical integrals, and there are $k^{2}=\frac{Q-P+6M}{2Q}$, $\sin^{2}\zeta_{\rm r}=\frac{Q-P+2M+4MP/r}{Q-P+6M}$, and $\sin^{2}\zeta_{\rm \infty}=\frac{Q-P+2M}{Q-P+6M}$. Thus, we can write $r$ as a function of $\alpha$ and $P$, i.e.,
\begin{equation}
\label{3-3}
\frac{1}{r} = \frac{P-2M-Q}{4MP} + \frac{Q-P+6M}{4MP} sn^{2}\Big(\frac{\gamma}{2}\sqrt{\frac{Q}{P}} + F(\zeta_{\rm \infty}, k)\Big).
\end{equation}
Thinking of this equation, the iso-radial curves for a given angle $\theta_{0}$ is derived. For the $(n+1)th$ order image of the KS BH accretion disk, the Eq. (\ref{3-2}) satisfies
\begin{equation}
\label{3-4}
2n \pi - \gamma = 2 \sqrt{\frac{P}{Q}} \Big(2K(k)- F(\zeta_{\rm r},k) - F(\zeta_{\rm \infty}, k)\Big),
\end{equation}
where $K(k)$ is the complete elliptic integral.

\begin{center}
\includegraphics[width=4.2cm,height=4.2cm]{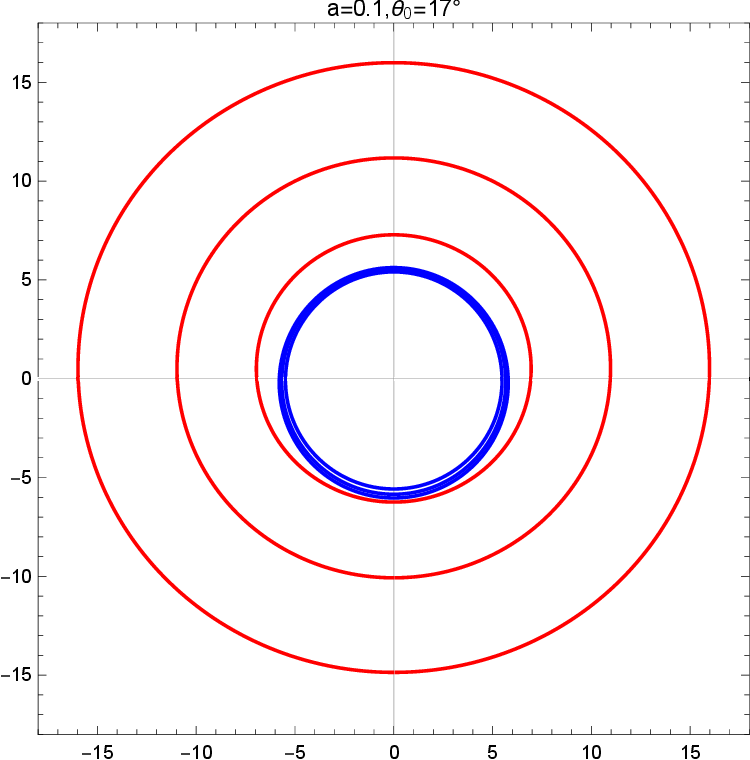}
  \hspace{0.5cm}
  \includegraphics[width=4.2cm,height=4.2cm]{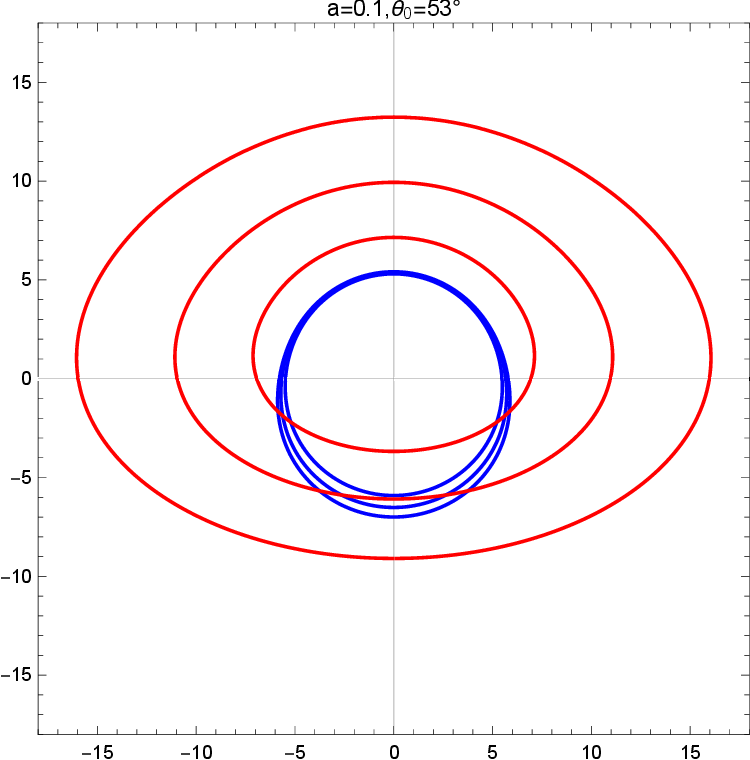}
  \hspace{0.5cm}
  \includegraphics[width=4.2cm,height=4.2cm]{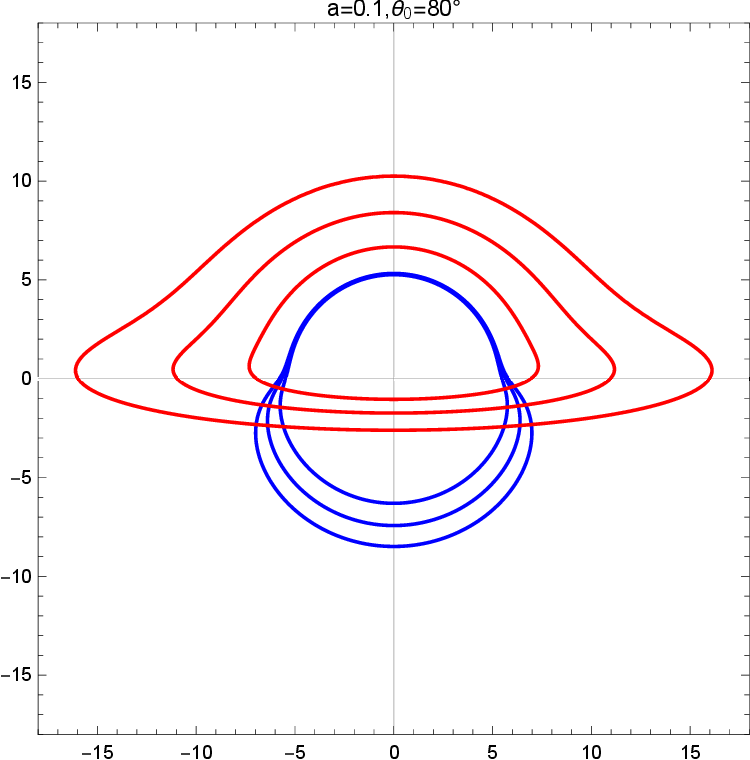}
  \hspace{0.5cm}
  \includegraphics[width=4.2cm,height=4.2cm]{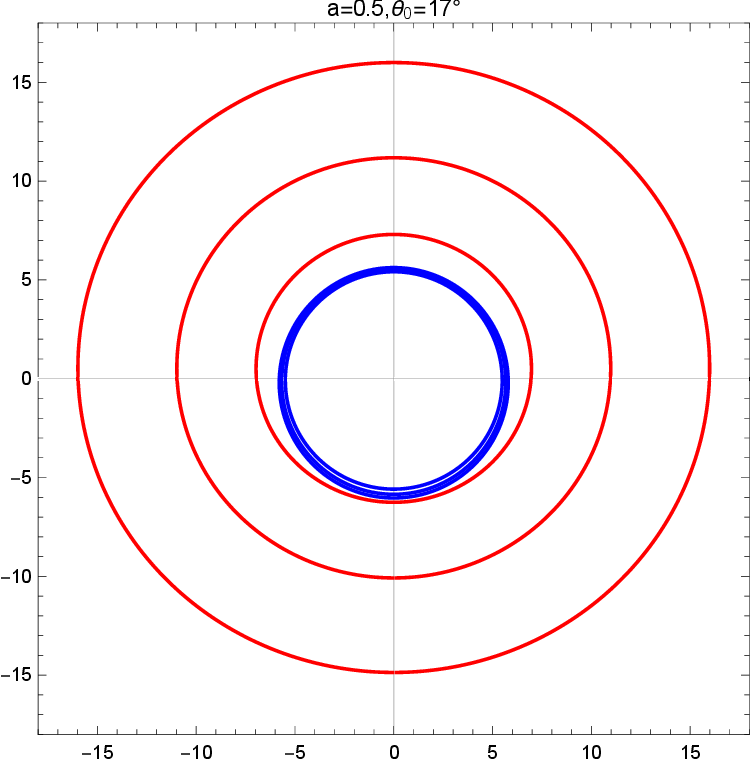}
  \hspace{0.5cm}
  \includegraphics[width=4.2cm,height=4.2cm]{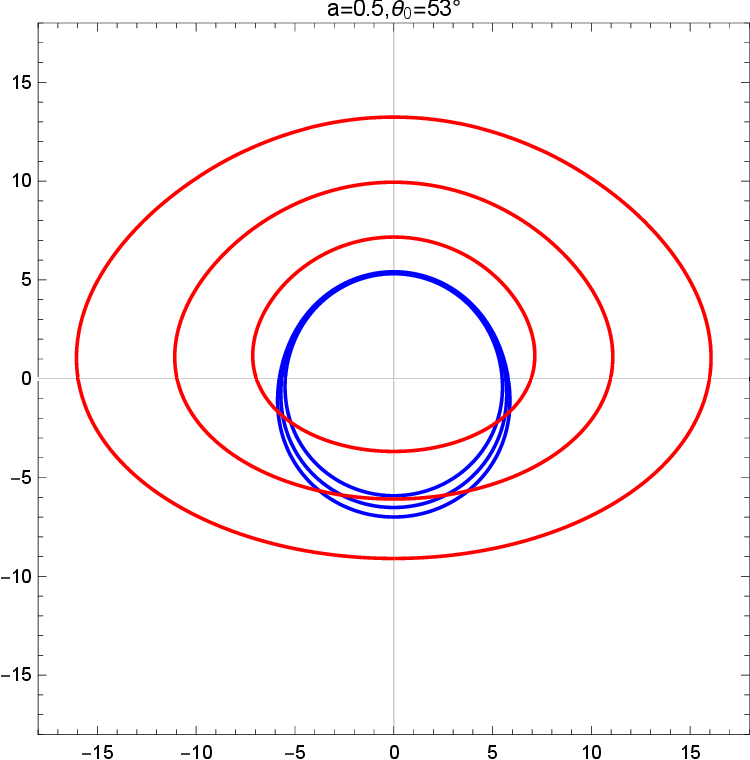}
  \hspace{0.5cm}
  \includegraphics[width=4.2cm,height=4.2cm]{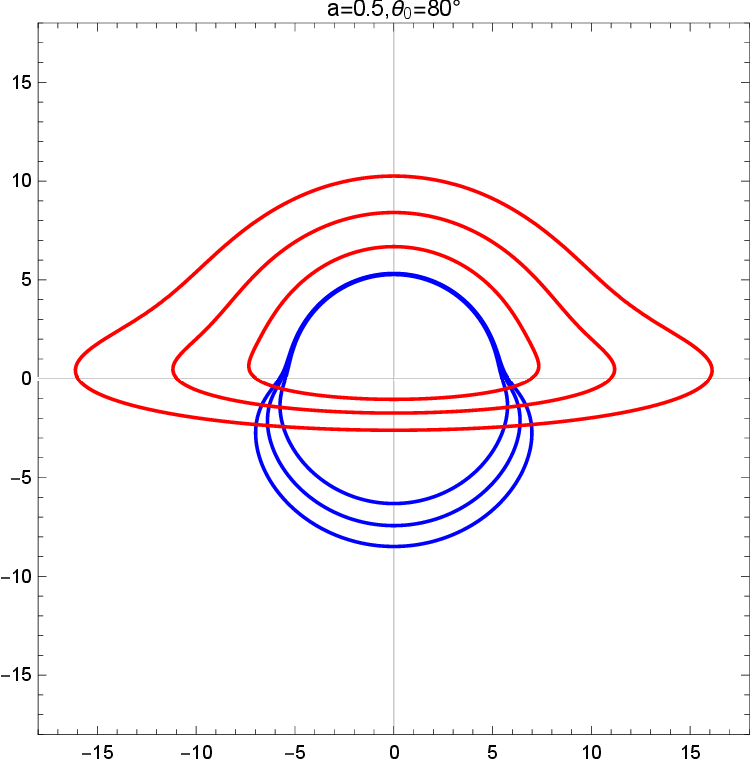}
\parbox[c]{15.0cm}{\footnotesize{\bf Fig~3.}  
The direct (red line) and secondary (blue line) images of KS BH accretion disk on the observation angles. The lines from inside out correspond to $r= 6.2M$, $r=10M$, and $r=15M$. The observation angles from left to right are $\theta_{0}=17^{\circ}$, $\theta_{0}=53^{\circ}$, and $\theta_{0}=80^{\circ}$, respectively. {\em Top Panel} -- the deformation parameter $a=0.1$ and {\em Bottom Panel} -- the deformation parameter $a = 0.5$. The BH mass is taken as $M = 1$.}
\label{fig3}
\end{center}

\par
Figure 3 depicts the direct and secondary images of photon orbits around the KS BH for different values of the deformation parameter. In the images presented, photons emitted in a direction above the equatorial plane are represented by the red line, while the blue lines correspond to secondary images generated by photons moving in a direction below the equatorial plane. It can be observed that the radius of the photon orbit decreases as the value of $a$ decreases. 
Moreover, our research results are in line with those of previous studies, indicating that the observation angle significantly influences the results obtained. This implies that the outcomes would be entirely different if viewed from different perspectives during observation.

\subsection{Radiation flux}
\label{sec:3-2}
The radiation flux of a thin accretion disk is discussed in this section. The expression of radiation flux is given in \cite{43,44}
\begin{equation}
\label{3-5}
F = - \frac{\dot{M}}{4\pi \sqrt{-g}} \frac{\Omega_{,\rm r}}{(E-\Omega L)^{2}} \int_{r_{\rm in}}^{r} (E- \Omega L)L_{,\rm r} {\rm d} r,
\end{equation}
where $\dot{M}$ is the mass accretion rate, $g$ is the metric determinant, and $r_{\rm in}$ represents the inner edge of the accretion disk. The $E$, $\Omega$ and $L$ denote the energy, angular momentum, and angular velocity, respectively. For a static spherically symmetric BH, the metric can be expressed as ${\rm d}s^{2}=g_{\rm tt} {\rm d}t^{2} + g_{\rm \phi\phi} {\rm d}\phi^{2} + g_{\rm rr}{\rm d}r^{2} + g_{\rm \theta\theta}{\rm d}\theta^{2}$. The $E$, $L$ and $\Omega$ are represented as
\begin{eqnarray}
\label{3-6}
&&E=-\frac{g_{\rm tt}}{\sqrt{-g_{\rm tt}-g_{\rm \phi\phi}\Omega^{2}}},\\
&&L=\frac{g_{\rm \phi\phi}\Omega}{\sqrt{-g_{\rm tt}-g_{\rm \phi\phi}\Omega^{2}}},\\
&&\Omega=\frac{{\rm d}\phi}{{\rm d}t}=\sqrt{-\frac{g_{\rm tt,r}}{g_{\rm \phi\phi,r}}}.
\end{eqnarray}
Utilizing the above equations, the radiant energy flux over the disk is obtained, we have
\begin{eqnarray}
\label{3-7}
&&F =\nonumber \\
&&\frac{-\dot{M} (6M h^{3}+4a^{2}r^{2}-3a^{4})}{4\sqrt{2}\pi r^{4} h^{2}(3a^{2}-2r^{2}+6M h)\sqrt{\frac{2Mr^{2}+a^{2}(-2M+h)}{r^{5}-a^{2}r^{3}}}}  \nonumber\\
&&\int_{r_{\rm in}}^{r} \frac{-2M(6a^{4}-6a^{2}r^{2}+r^{4})+12M^{2}h^{3}+3a^{4}h}{r^{2}h^{2}(3a^{2}-2r^{2}+6Mh)\sqrt{\frac{4Mr^{2}+2a^{2}(-2M+h)}{r^{5}-a^{2}r^{3}}}} {\rm d} r,
\end{eqnarray}
where $h \equiv \sqrt{r^{2}-a^{2}}$.

\begin{center}
\includegraphics[width=4.2cm,height=3.6cm]{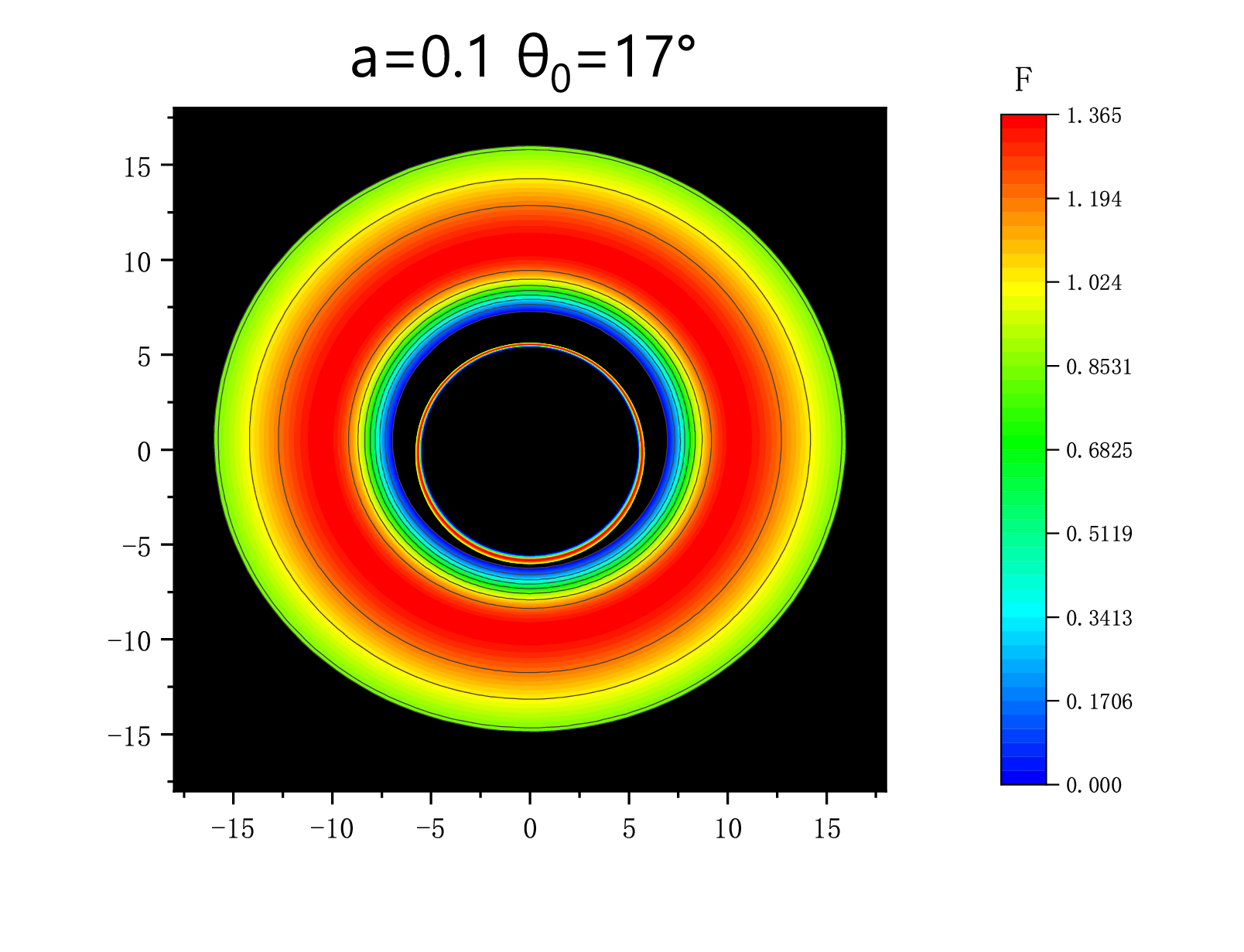}
  \hspace{0.5cm}
  \includegraphics[width=4.2cm,height=3.6cm]{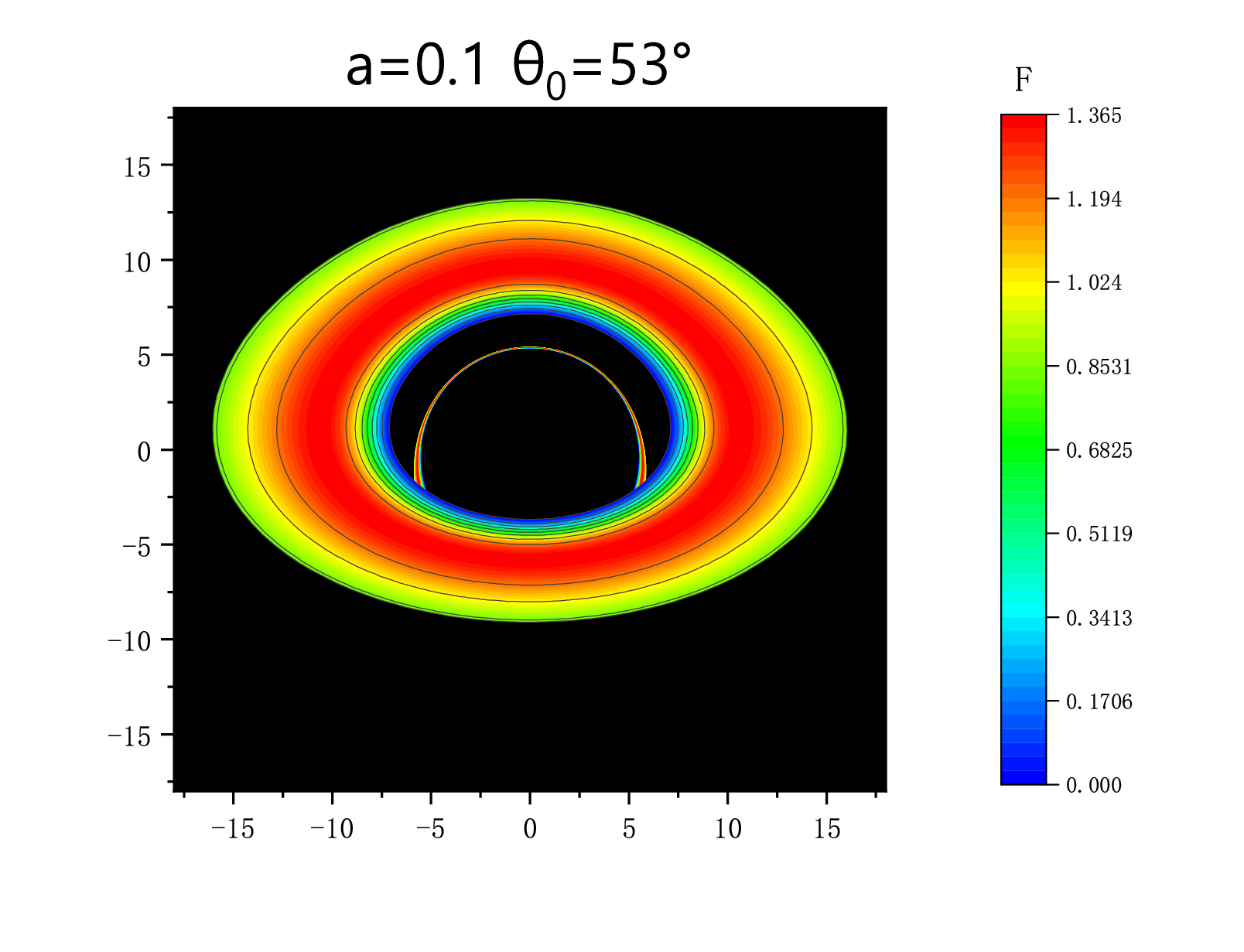}
  \hspace{0.5cm}
  \includegraphics[width=4.2cm,height=3.6cm]{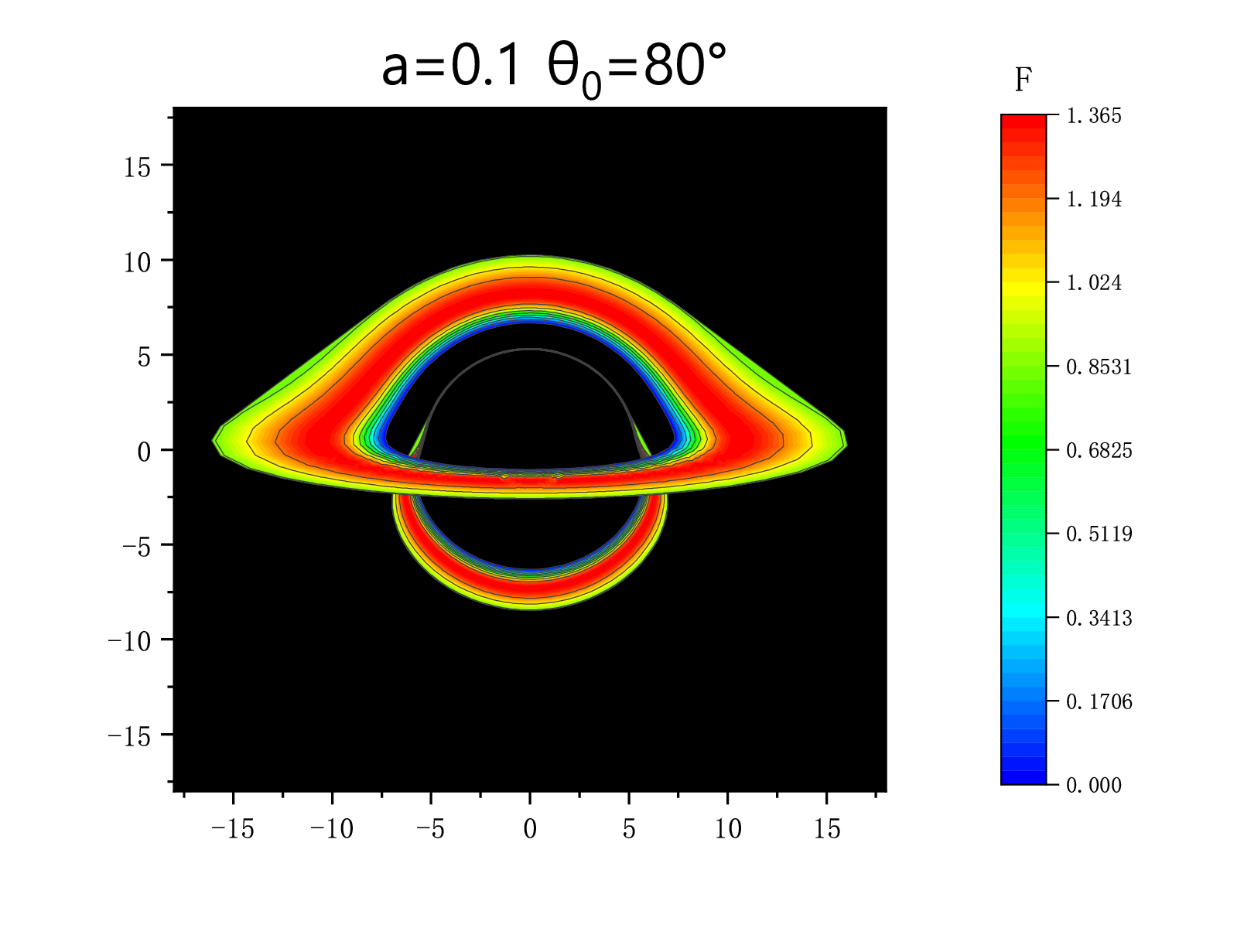}
  \hspace{0.5cm}
  \includegraphics[width=4.2cm,height=3.6cm]{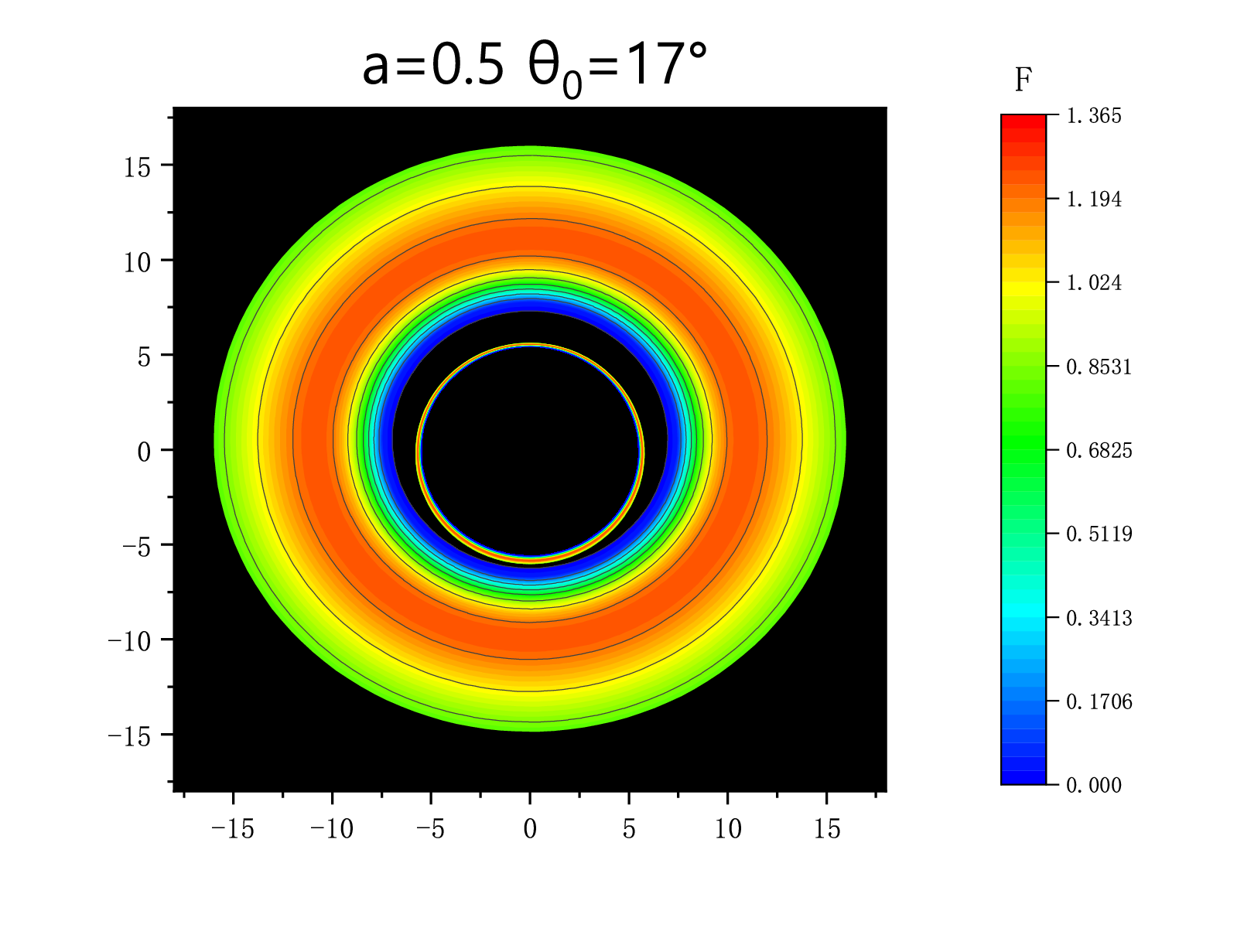}
  \hspace{0.5cm}
  \includegraphics[width=4.2cm,height=3.6cm]{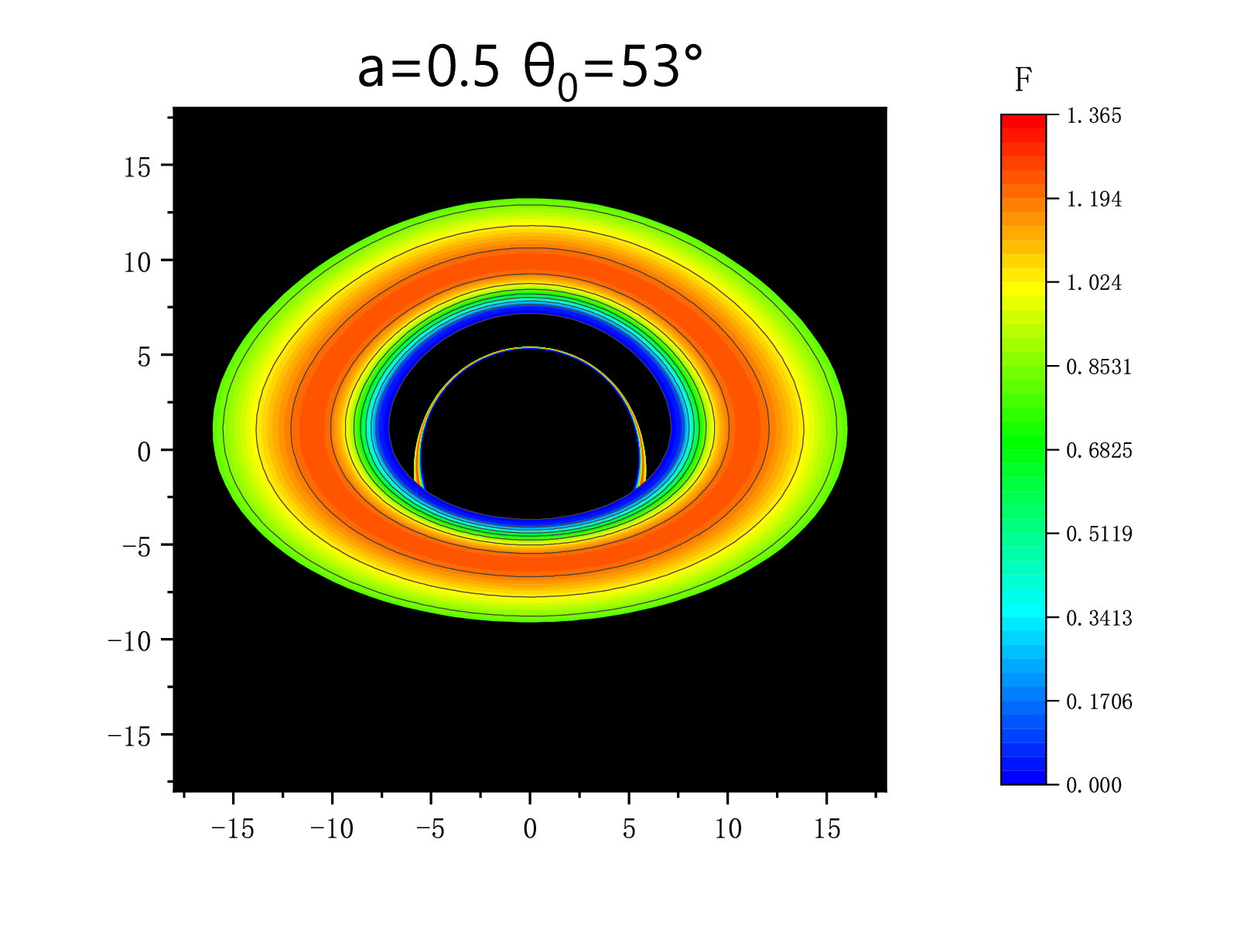}
  \hspace{0.5cm}
  \includegraphics[width=4.2cm,height=3.6cm]{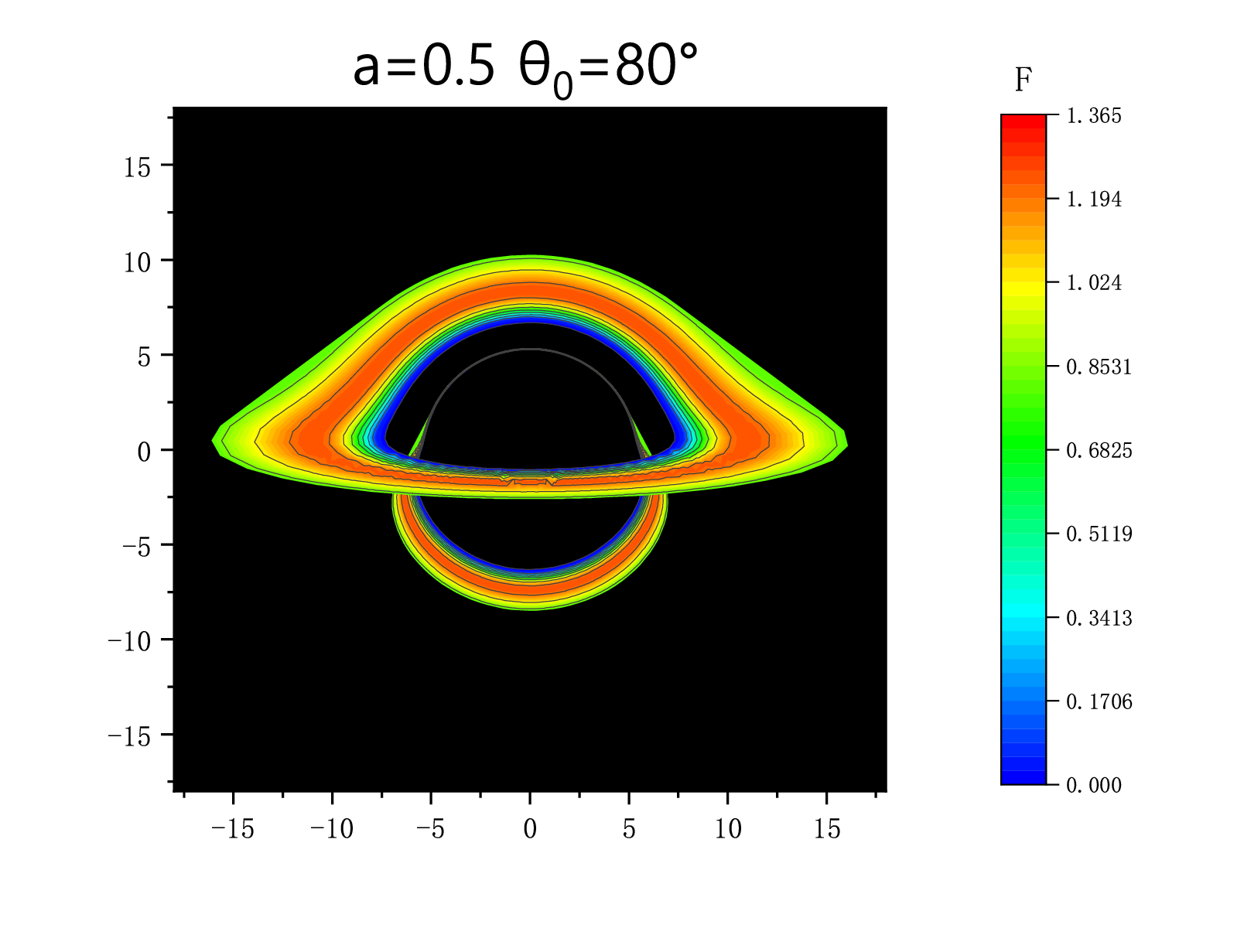}
\parbox[c]{15.0cm}{\footnotesize{\bf Fig~4.}  
The intrinsic flux of the source $F$ of the accretion disk of the KS BH is demonstrated at the inclination angles of the observer $\theta_{0}=17^{\circ}$, $53^{\circ}$, and $80^{\circ}$. The inner edge of the disk at $r_{in} = r_{isco}$, and the outer edge of the disk is at $r= 15M$. {\em Top Panel}: the deformation parameter $a = 0.1$. {\em Bottom Panel}: the deformation parameter $a = 0.5$. The BH mass is taken as $M = 1$.}
\label{fig4}
\end{center}

\par
Figure 4 depicts the radiation intensity of the accretion disk. It is evident that an increase in the deformation parameter $a$ causes a notable decrease in the flux ($F$) for large values. The main reason for this phenomenon is that the strength of the gravitational field decreases with the increase of $a$, which leads to the change of the motion speed of the matter on the accretion disk around the BH, resulting in the change of temperature and other properties, and ultimately the reduction of radiant flux. Additionally, the figure shows that with an increase in the tilt angle, the separation between the direct and secondary images becomes more pronounced, and the direct image takes on a more prominent ``hat-like'' shape. In order to visually demonstrate the relationship between radiation flux and deformation parameter, we assume three functional models. These are \emph{Case 1:} Fourier function, \emph{Case 2:} Exponential function, and \emph{Case 3:} Multinomial function. The red dots represent the accurate values of $F$ as a function of $a$. As $a$ increases, the value of $F$ decreases, while the slope of the curve increases with the increase of $a$. Additionally, we fit this curve using three hypothetical functional models and find that the exponential function, under certain fixed parameters, is more consistent with the results (see Fig. 5).

\begin{center}
 \includegraphics[width=6.5cm,height=5cm]{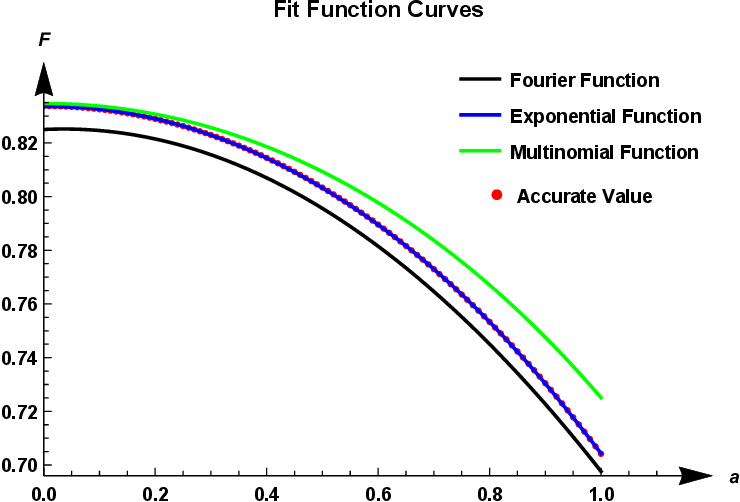}
\parbox[c]{15.0cm}{\footnotesize{\bf Fig~5.}  
Fitting of fourier function (the black solid line), exponential function(the bule solid line), and multinomial function (the green solid line) to the accurate value of the intrinsic flux of the source $F$ and the deformation parameter $a$.}
\label{fig5}
\end{center}

\section{Observable features of the thin disk}
\label{sec:4}
\subsection{Observed flux}
\label{sec:4-2}
\par
 The gravity of a black hole affects matter on the accretion disk at different distances, resulting in varying degrees of gravitational red shift. Furthermore, due to the rotation of the disk, matter on the plane perpendicular to the rotation axis experiences different rates and directions of motion, leading to different degrees of red shift or blue shift. In the left half of the disk, the contribution of Doppler blueshift caused by the disk's rotation can exceed the gravitational red shift, resulting in a blueshift for the corresponding photons. In this scenario, we can calculate the observed flux $F_{\rm obs}$ of the KS BH surrounding the accretion disk. Following the Ref. \cite{5}, the observed flux $F_{\rm obs}$ satisfies
\begin{equation}
\label{4-1}
F_{\rm obs} = \frac{F}{(1+z)^{4}}.
\end{equation}
When a photon is discharged from a source particle in orbit around a BH, its energy can be expressed as the dot product of the photon's 4-momentum vector $p$ and the 4-velocity vector $\mu$ of the emitting particle. We have
\begin{equation}
\label{4-2}
E_{\rm em} = p_{\rm t} \mu^{\rm t} + p_{\rm \phi} \mu^{\rm \phi} = p_{\rm t} \mu^{\rm t} \Bigg(1 + \Omega \frac{p_{\rm \phi}}{p_{\rm t}}\Bigg),
\end{equation}
where $p_{\rm t}$ and $p_{\rm \phi}$ represent the photon 4-momentum. For a distant observer, $p_{\rm t} / p_{\rm \phi}$ is nothing but the impact parameter of the photon relative the z-axis. According Eq. (\ref{3-1}) and trigonometric function relationship
\begin{equation}
\label{4-3}
\sin \theta_{0} \cos \alpha = \cos \gamma  \sin \beta,
\end{equation}
one can obtain
\begin{equation}
\label{4-4}
\frac{p_{\rm t}}{p_{\rm \phi}} = b \sin \theta_{0} \sin \alpha.
\end{equation}
Thus, the redshift factor is
\begin{equation}
\label{4-5}
1 + z = \frac{E_{\rm em}}{E_{\rm obs}} = \frac{1 + b \Omega \cos \beta}{\sqrt{-g_{\rm tt} - 2 g_{\rm t \phi} - g_{\rm \phi\phi}}}.
\end{equation}
Using the KS BH metric, the above equation can be re-written as
\begin{equation}
\label{4-6}
1+z = \frac{2 + b \sqrt{\frac{4Mr^{2}+2a^{2}(-2M+\sqrt{r^{2}-a^{2}})}{r^{5}-a^{2}r^{3}}} \sin \alpha \sin \theta_{0}}{\sqrt{\frac{4r^{2}(\sqrt{r^{2}-a^{2}}-3M)-6a^{2}(\sqrt{r^{2}-a^{2}}-2M)}{r^{3}-a^{2}r}}}.
\end{equation}

\begin{center}
\includegraphics[width=4.2cm,height=3.6cm]{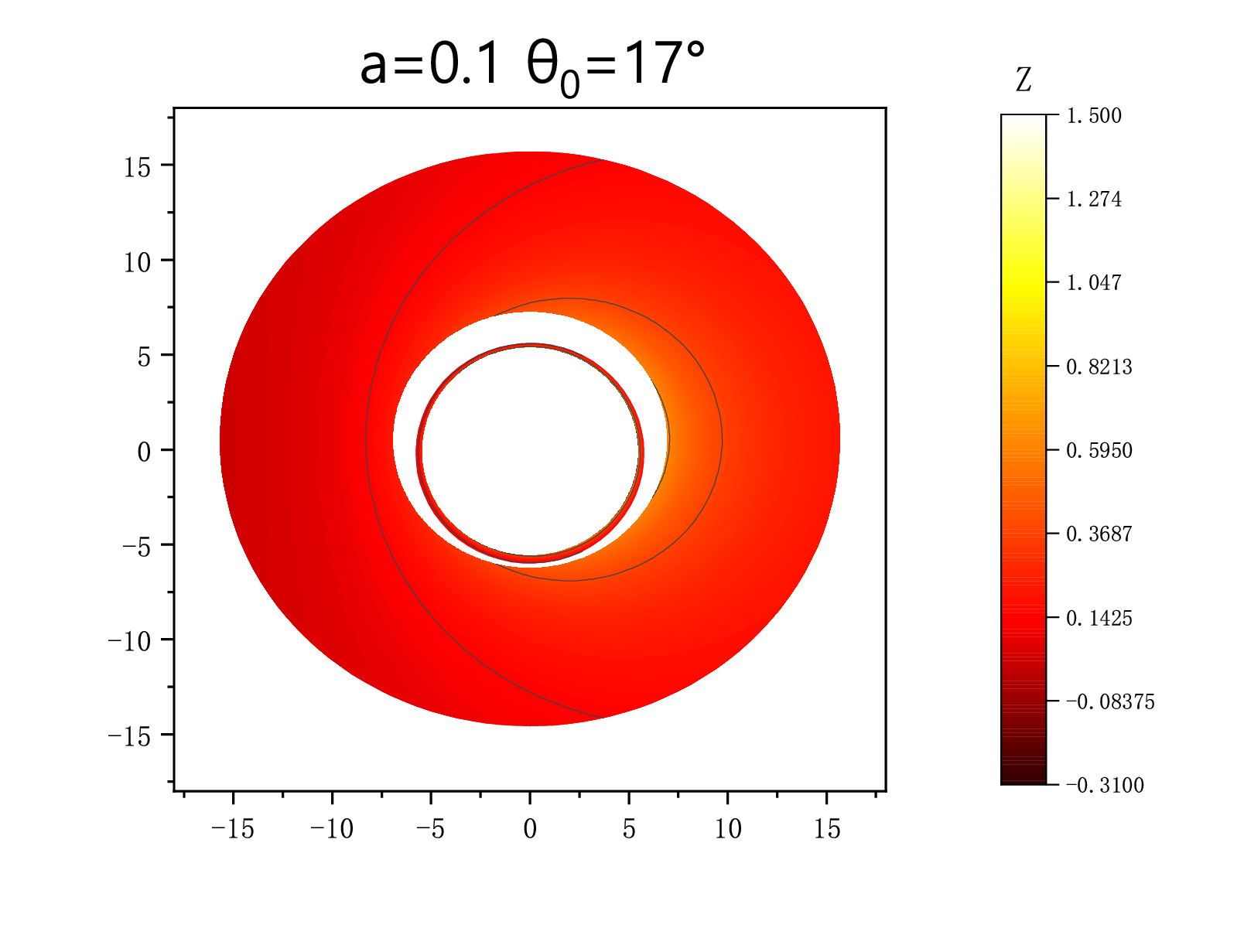}
  \hspace{0.5cm}
  \includegraphics[width=4.2cm,height=3.6cm]{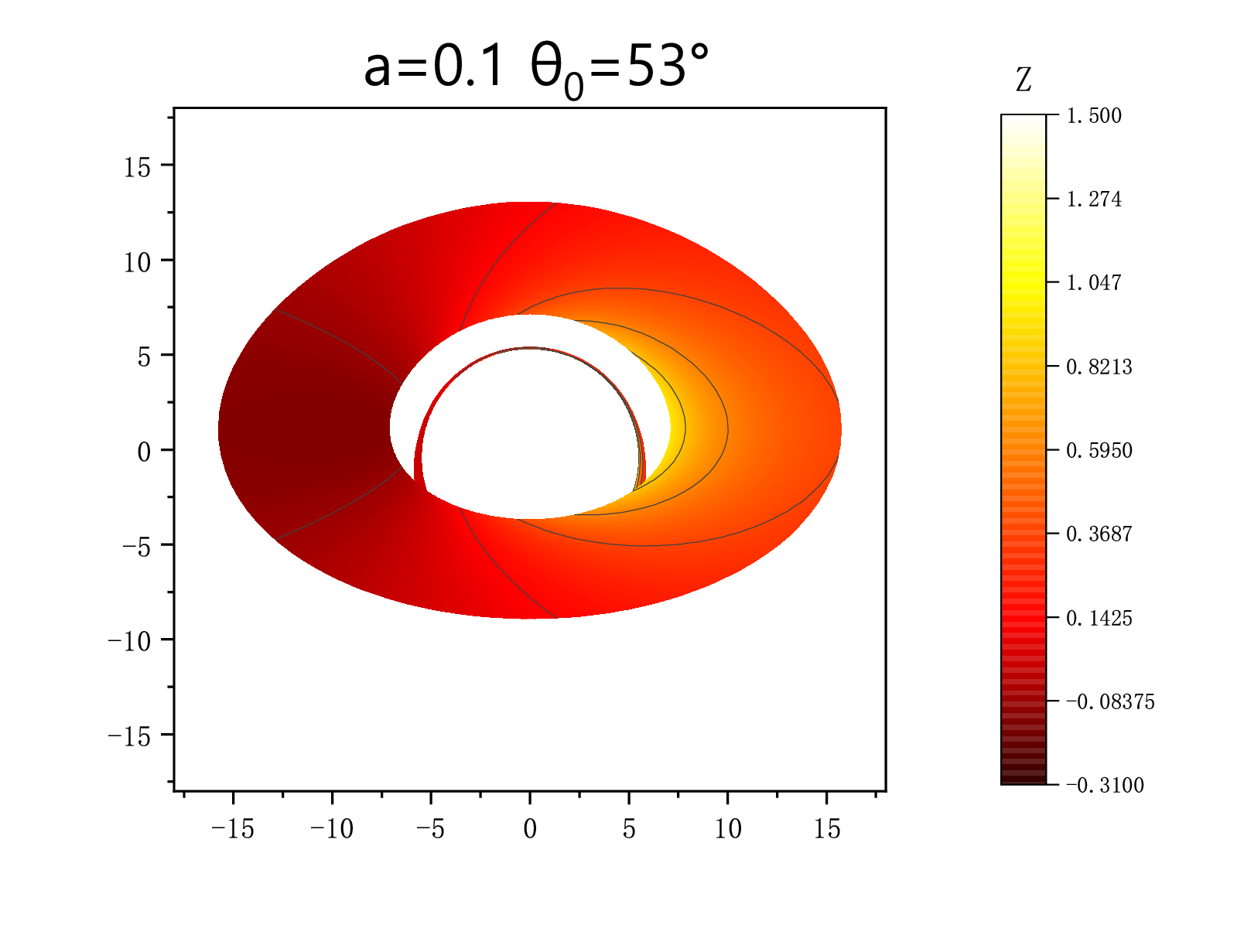}
  \hspace{0.5cm}
  \includegraphics[width=4.2cm,height=3.6cm]{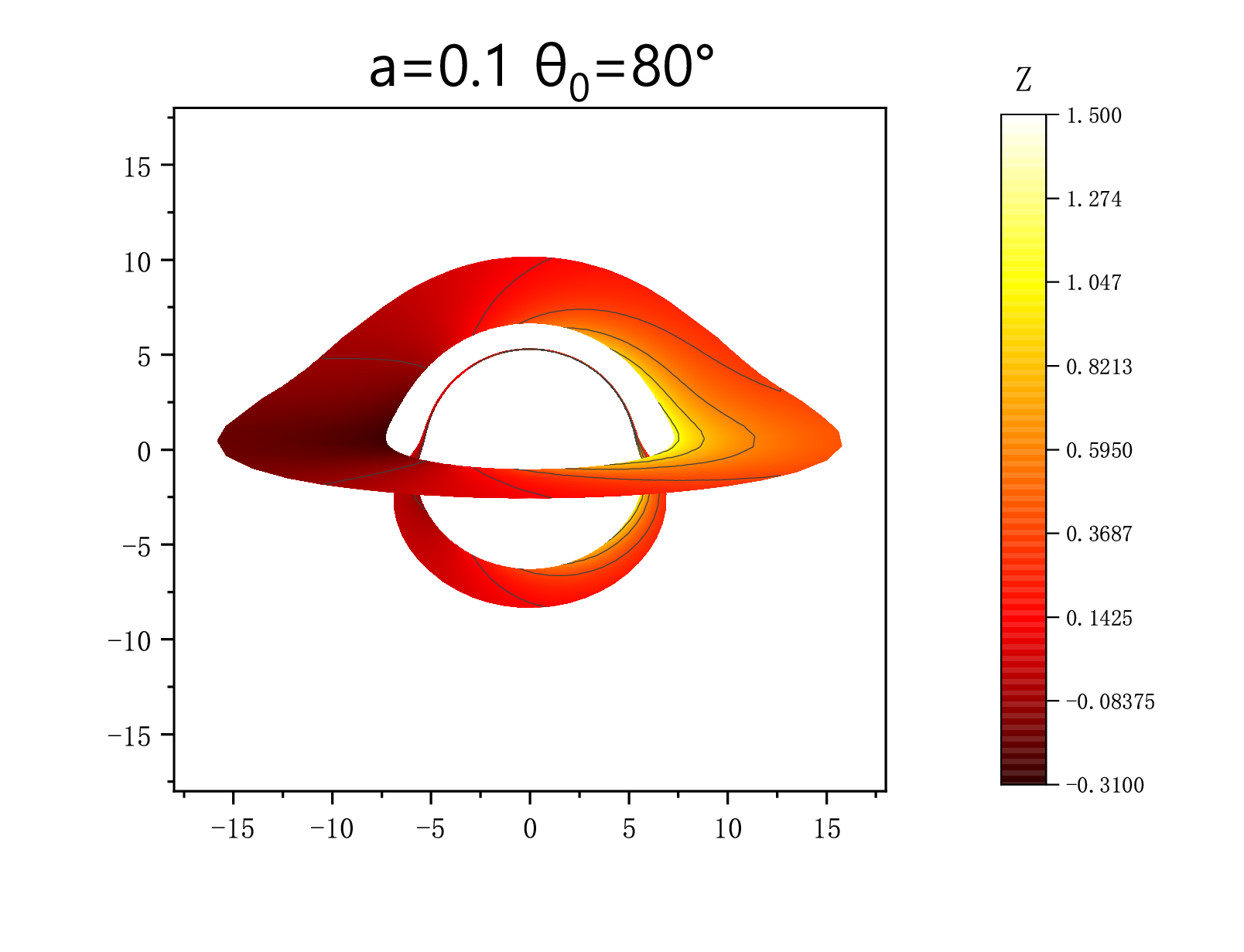}
  \hspace{0.5cm}
  \includegraphics[width=4.2cm,height=3.6cm]{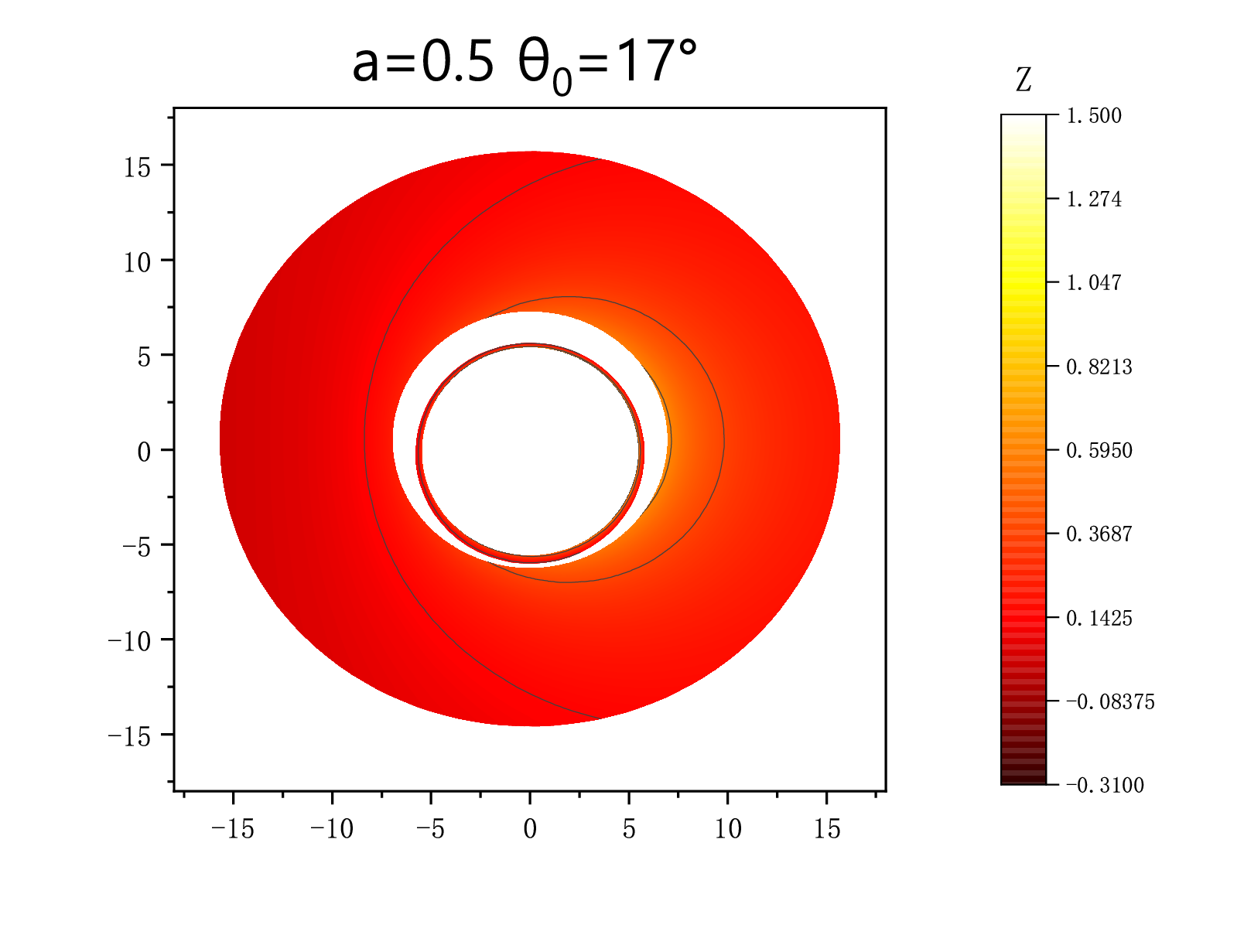}
  \hspace{0.5cm}
  \includegraphics[width=4.2cm,height=3.6cm]{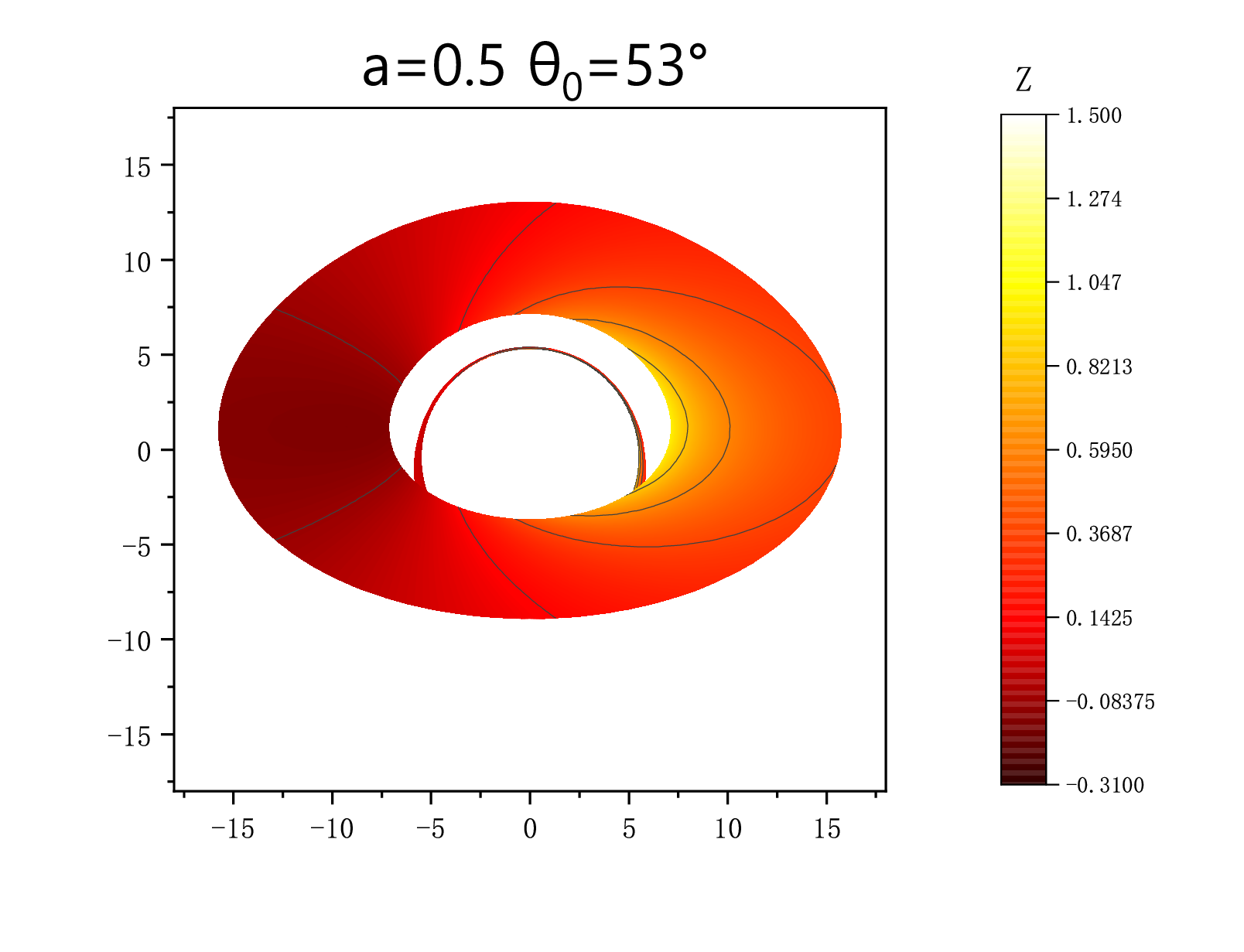}
  \hspace{0.5cm}
  \includegraphics[width=4.2cm,height=3.6cm]{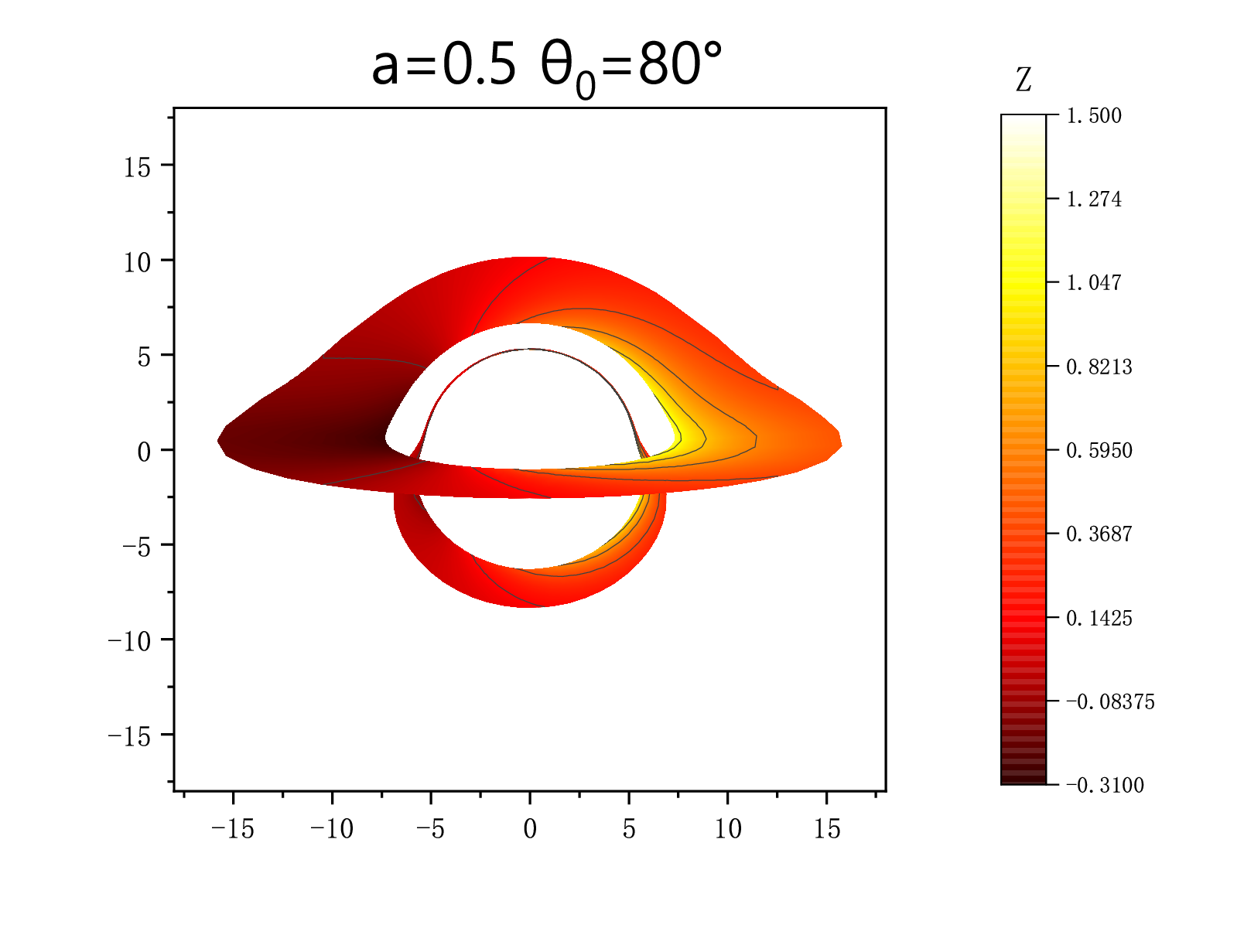}
\parbox[c]{15.0cm}{\footnotesize{\bf Fig~6.}  
Redshift distribution (curves of constant redshift $z$) of the accretion disk of the KS BH at the inclination angles of the observer $\theta_{0}=17^{\circ}$, $53^{\circ}$, and $80^{\circ}$. The inner edge of the disk at $r_{in} = r_{isco}$, and the outer edge of the disk is at $r= 15M$. {\em Top Panel}: the deformation parameter is $a = 0.1$ and {\em Bottom Panel}: the deformation parameter is $a = 0.5$. The BH mass is taken as $M = 1$.}
\label{fig6}
\end{center}

\par
Figure 6 illustrates the redshift images of the accretion disk with different parameter values. It is apparent that both red and blue shifts are present simultaneously. Moreover, as the observation angle $\theta_{0}$ increases gradually, the range of the redshift also gradually expands, which is consistent with the findings reported in \cite{11}. Due to the small impact of the change in the deformation parameter $a$ on the redshift $z$, the change in the redshift image in the figure is not significant.
\begin{center}
\includegraphics[width=4.2cm,height=3.6cm]{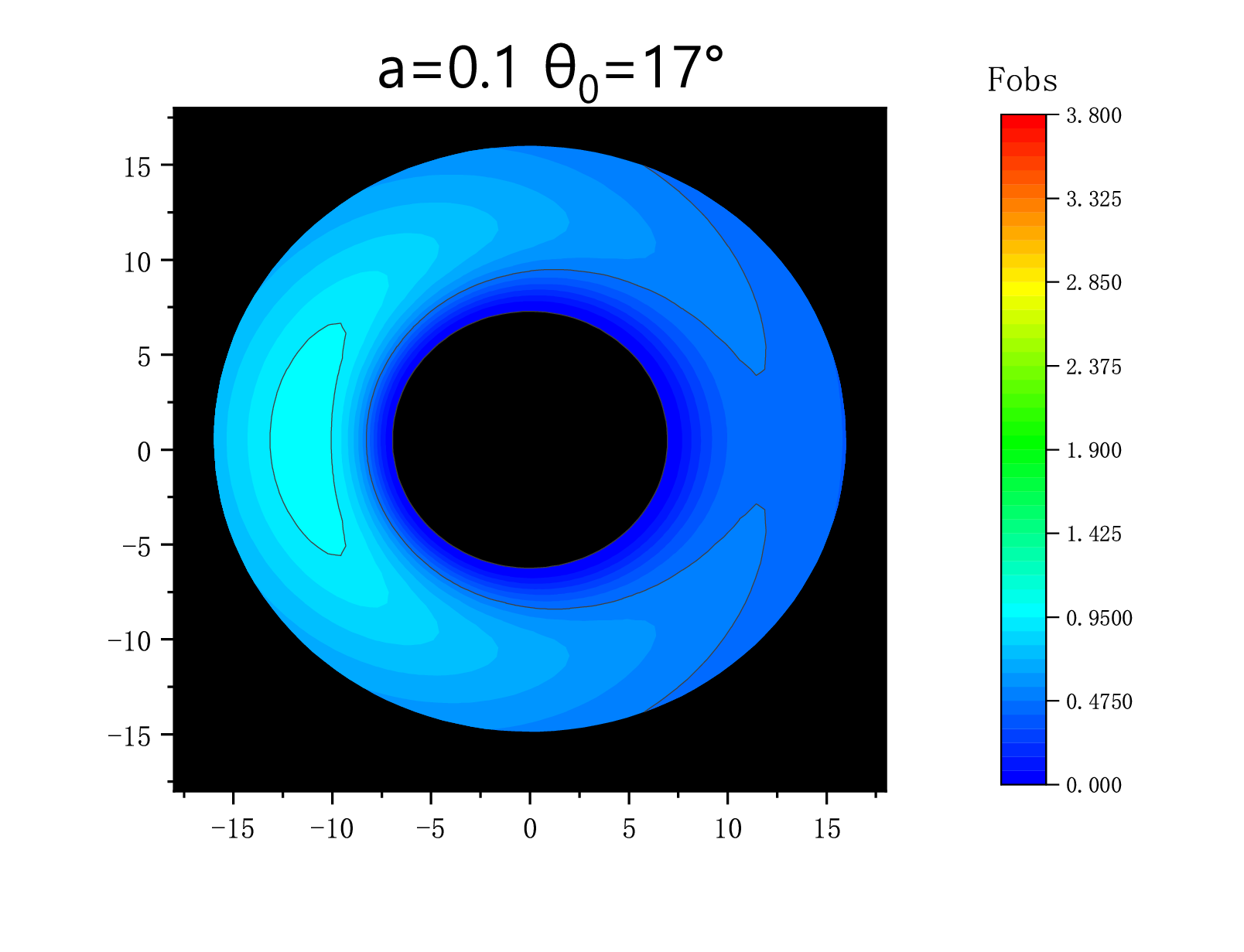}
  \hspace{0.5cm}
  \includegraphics[width=4.2cm,height=3.6cm]{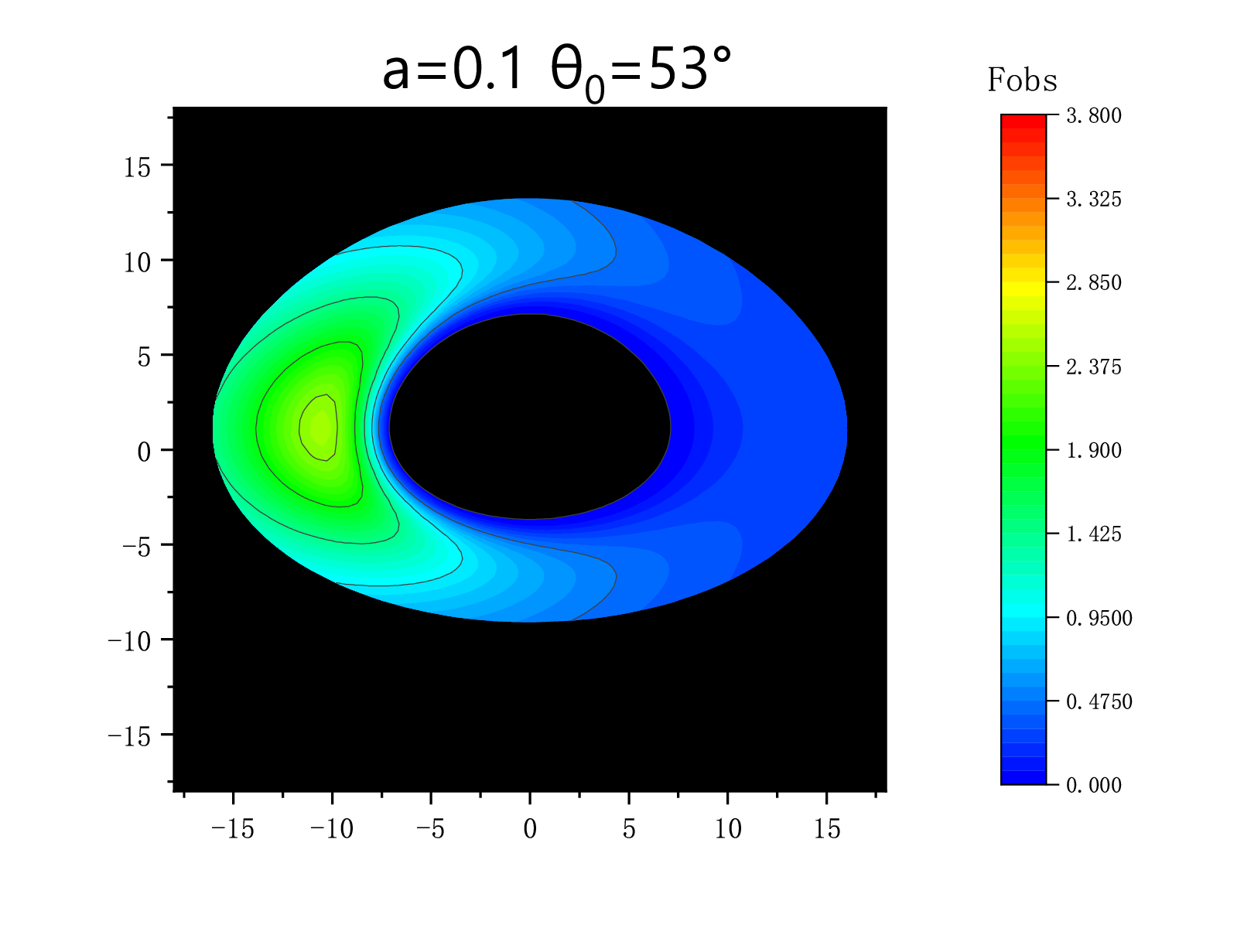}
  \hspace{0.5cm}
  \includegraphics[width=4.2cm,height=3.6cm]{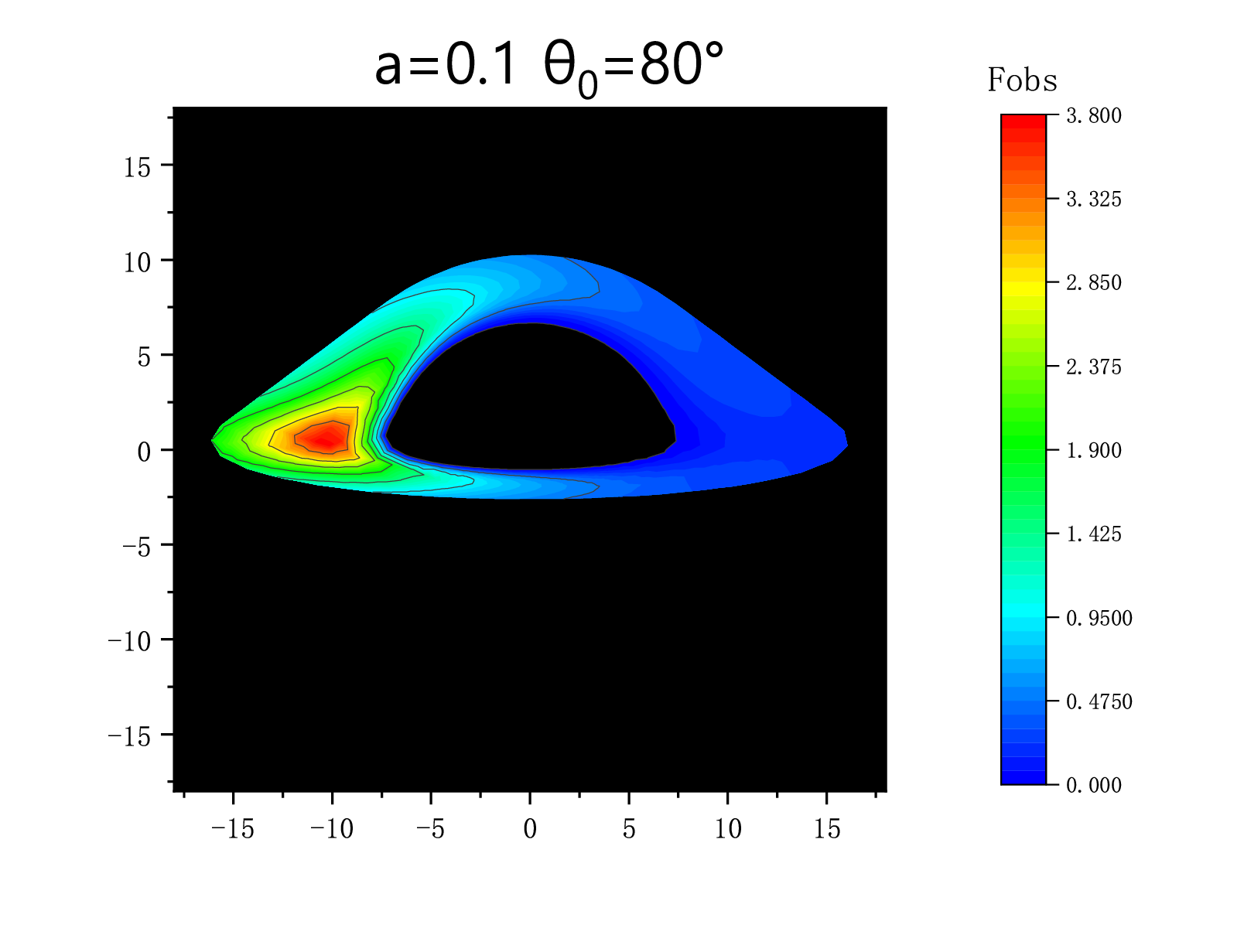}
  \hspace{0.5cm}
  \includegraphics[width=4.2cm,height=3.6cm]{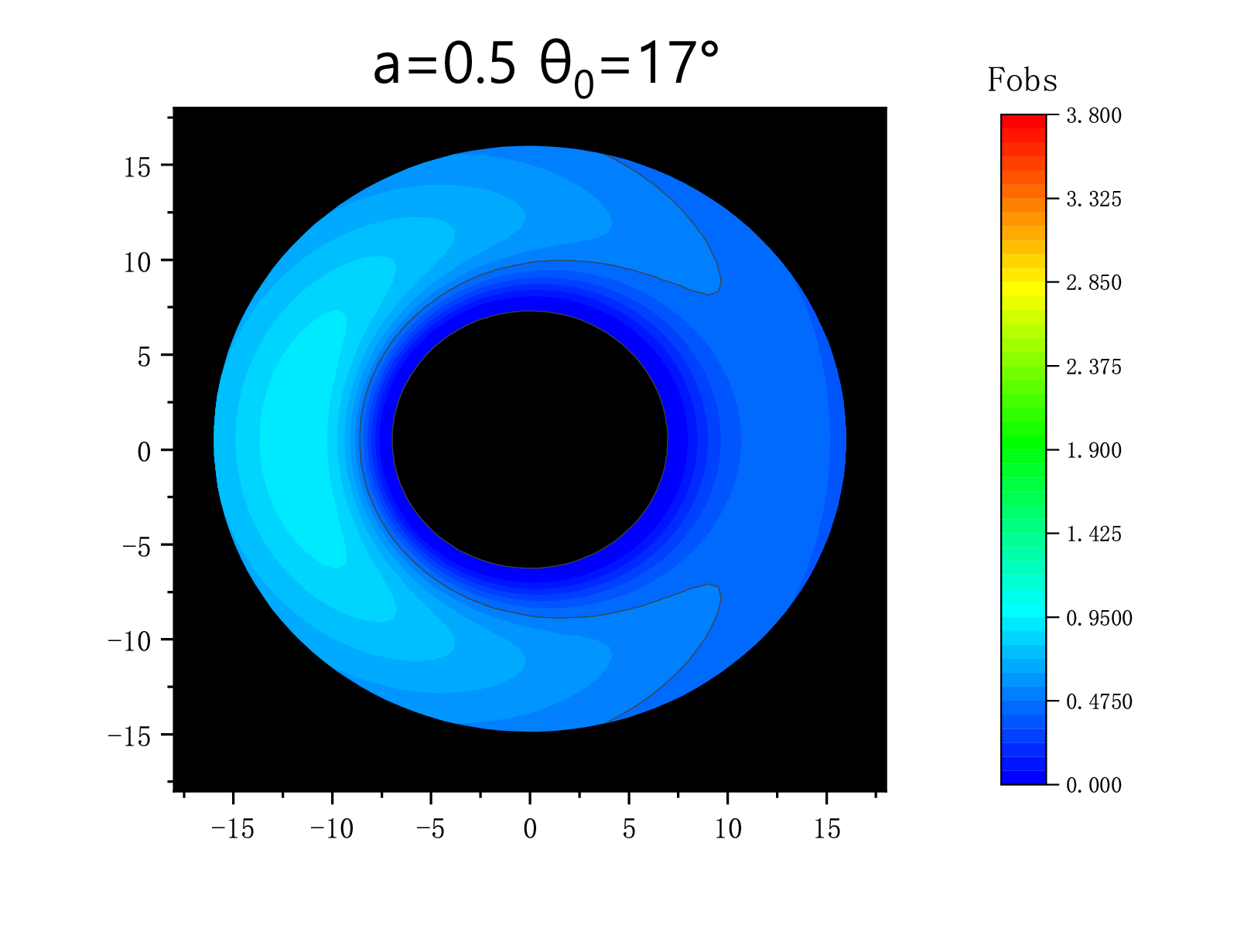}
  \hspace{0.5cm}
  \includegraphics[width=4.2cm,height=3.6cm]{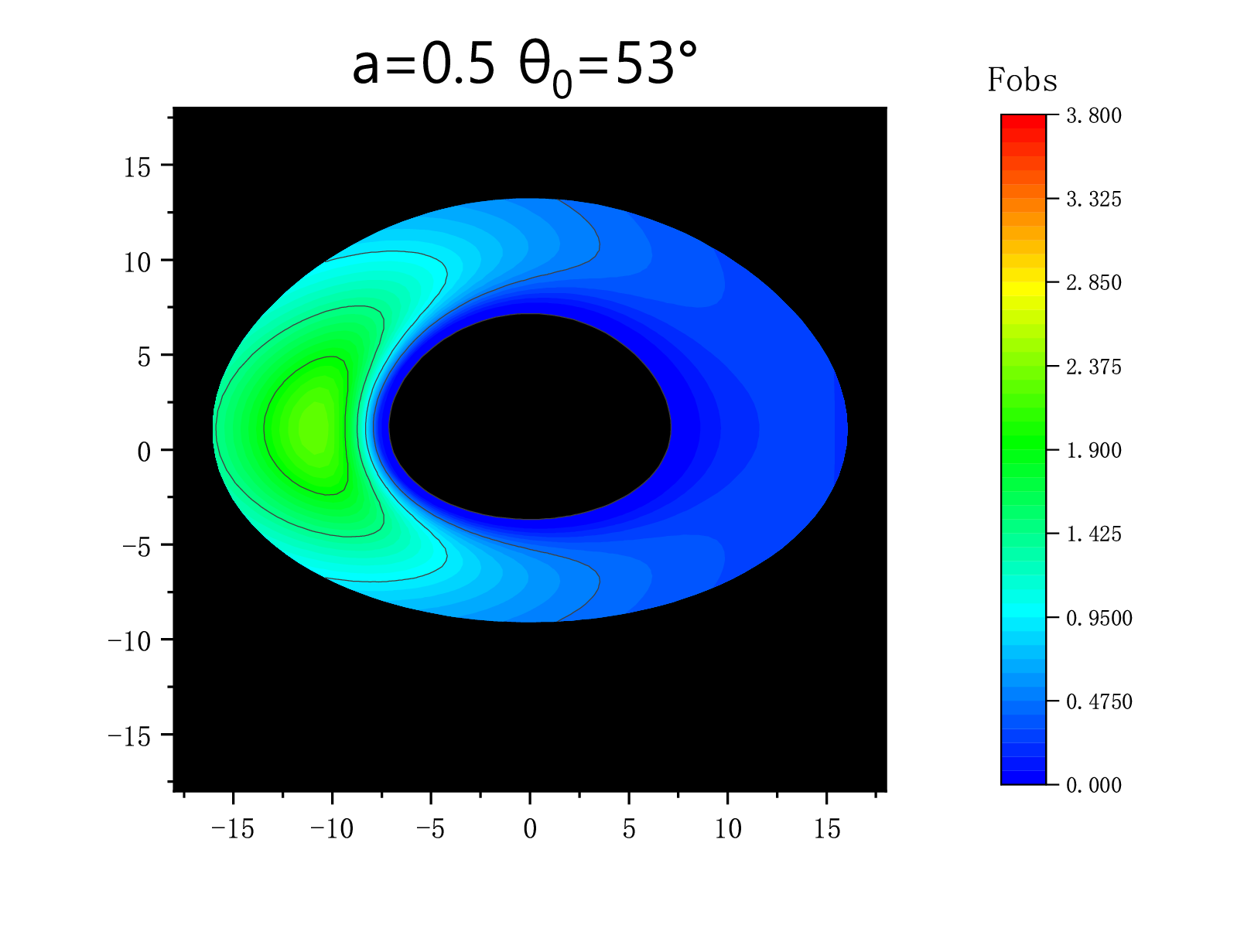}
  \hspace{0.5cm}
  \includegraphics[width=4.2cm,height=3.6cm]{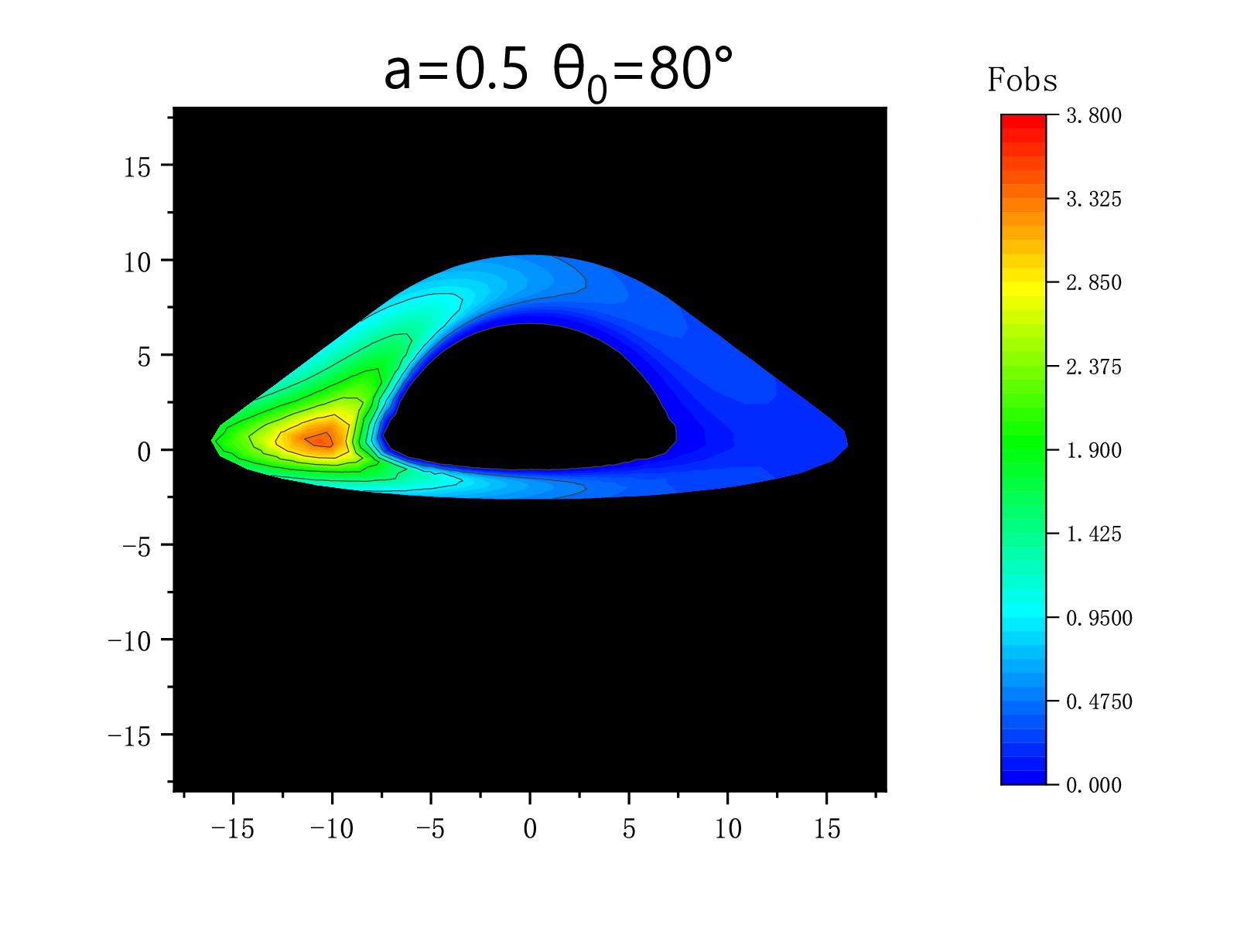}
\parbox[c]{15.0cm}{\footnotesize{\bf Fig~7.}  
The observed flux $F_{\rm obs}$ of the KS direct image with the inclination angles of the observer $\theta_{0}=17^{\circ}$, $53^{\circ}$, and $80^{\circ}$. The inner edge of the disk at $r_{\rm in} = r_{isco}$, and the outer edge of the disk is at $r= 15M$. {\em Top Panel}: the deformation parameter $a = 0.1$ and {\em Bottom Panel}: the deformation parameter $a = 0.5$. The BH mass is taken as $M = 1$.}
\label{fig7}
\end{center}

\begin{center}
\includegraphics[width=4.2cm,height=3.6cm]{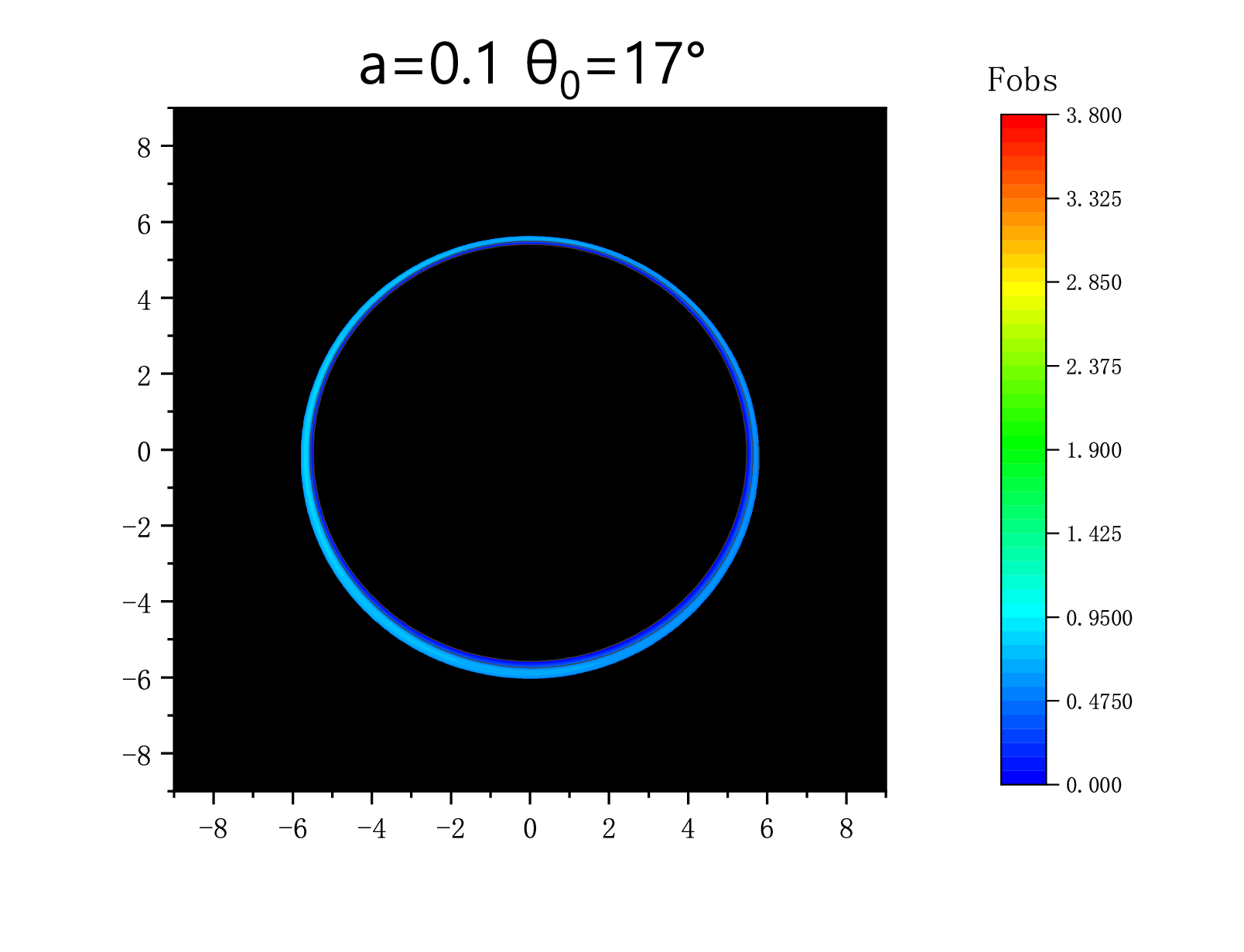}
  \hspace{0.5cm}
  \includegraphics[width=4.2cm,height=3.6cm]{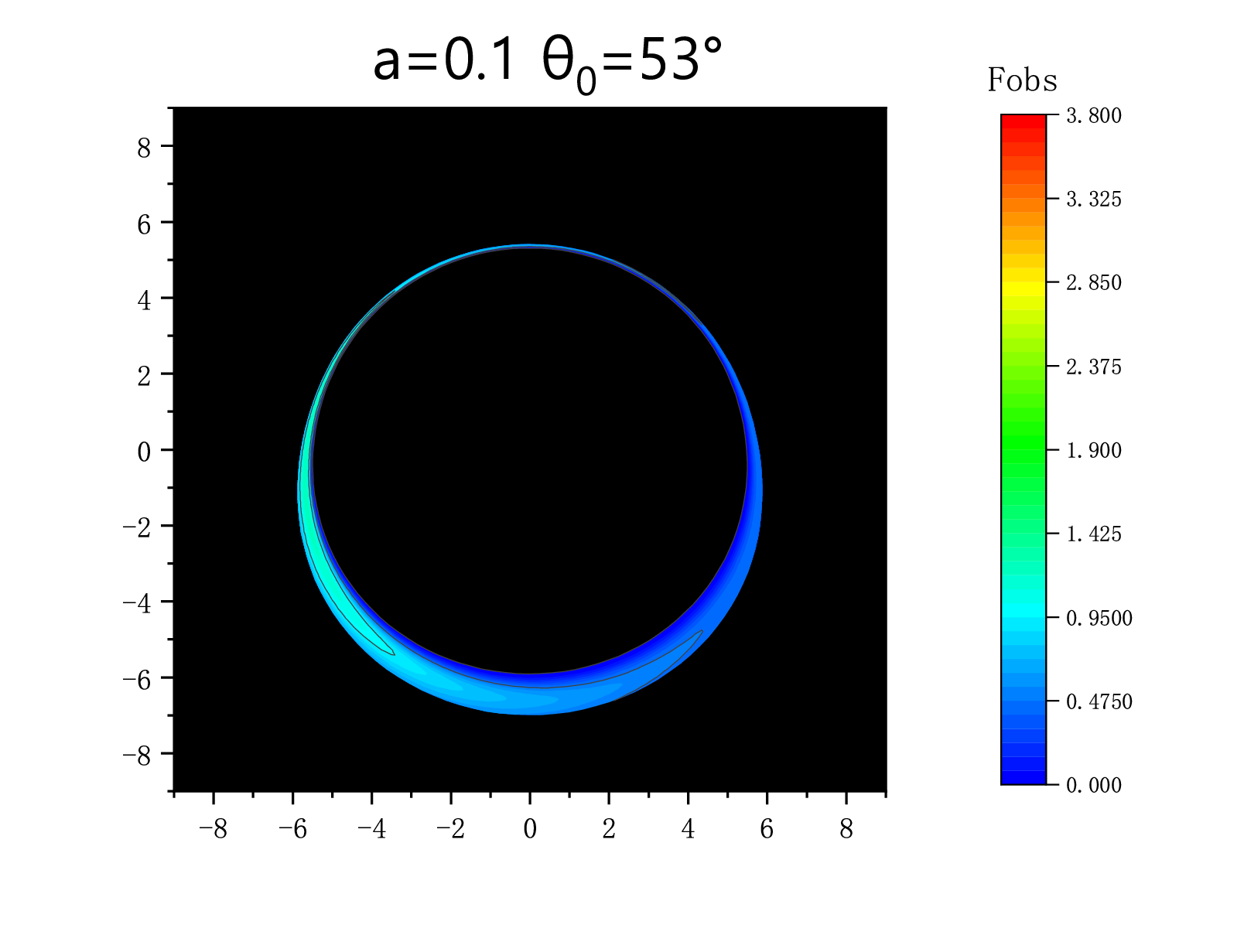}
  \hspace{0.5cm}
  \includegraphics[width=4.2cm,height=3.6cm]{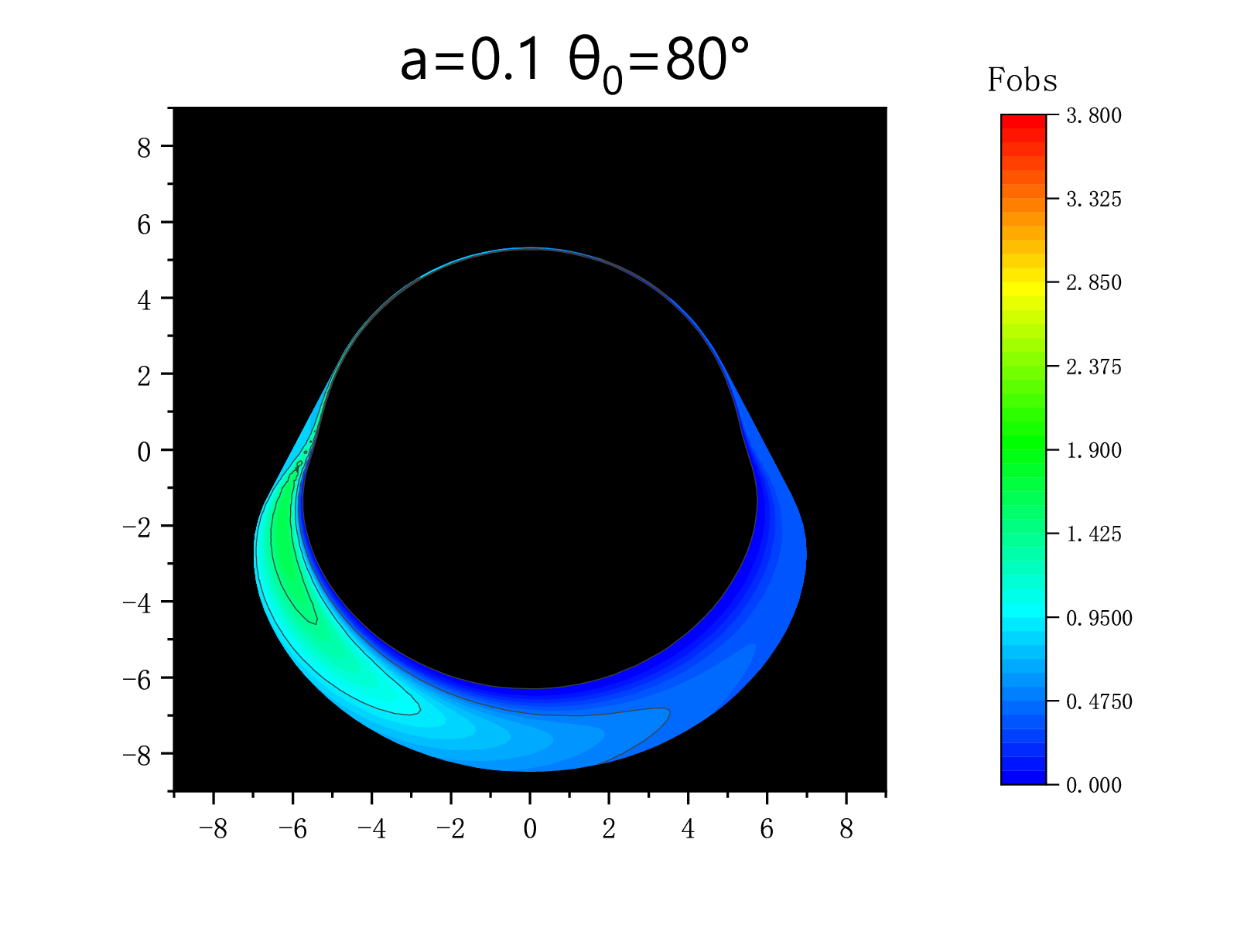}
  \hspace{0.5cm}
  \includegraphics[width=4.2cm,height=3.6cm]{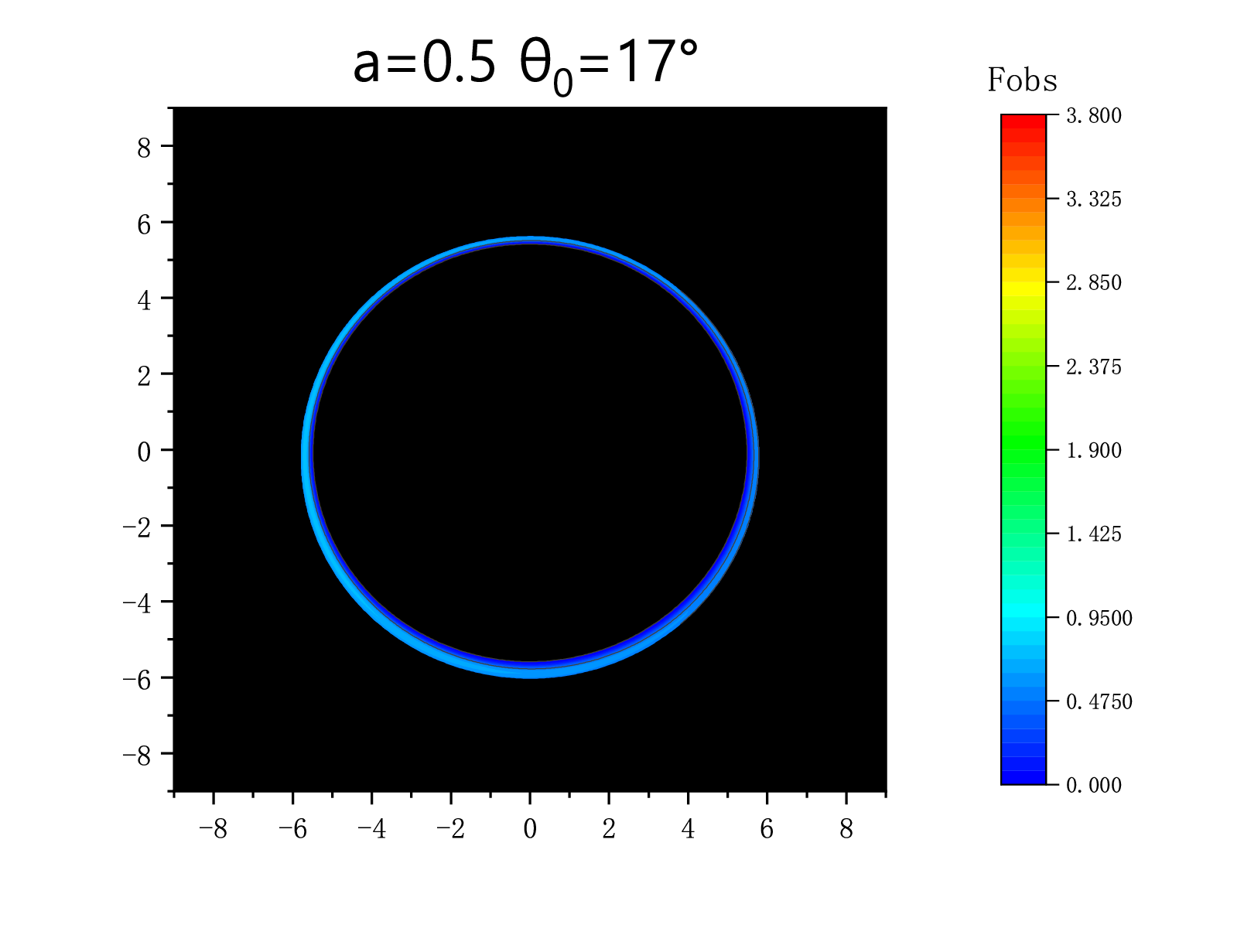}
  \hspace{0.5cm}
  \includegraphics[width=4.2cm,height=3.6cm]{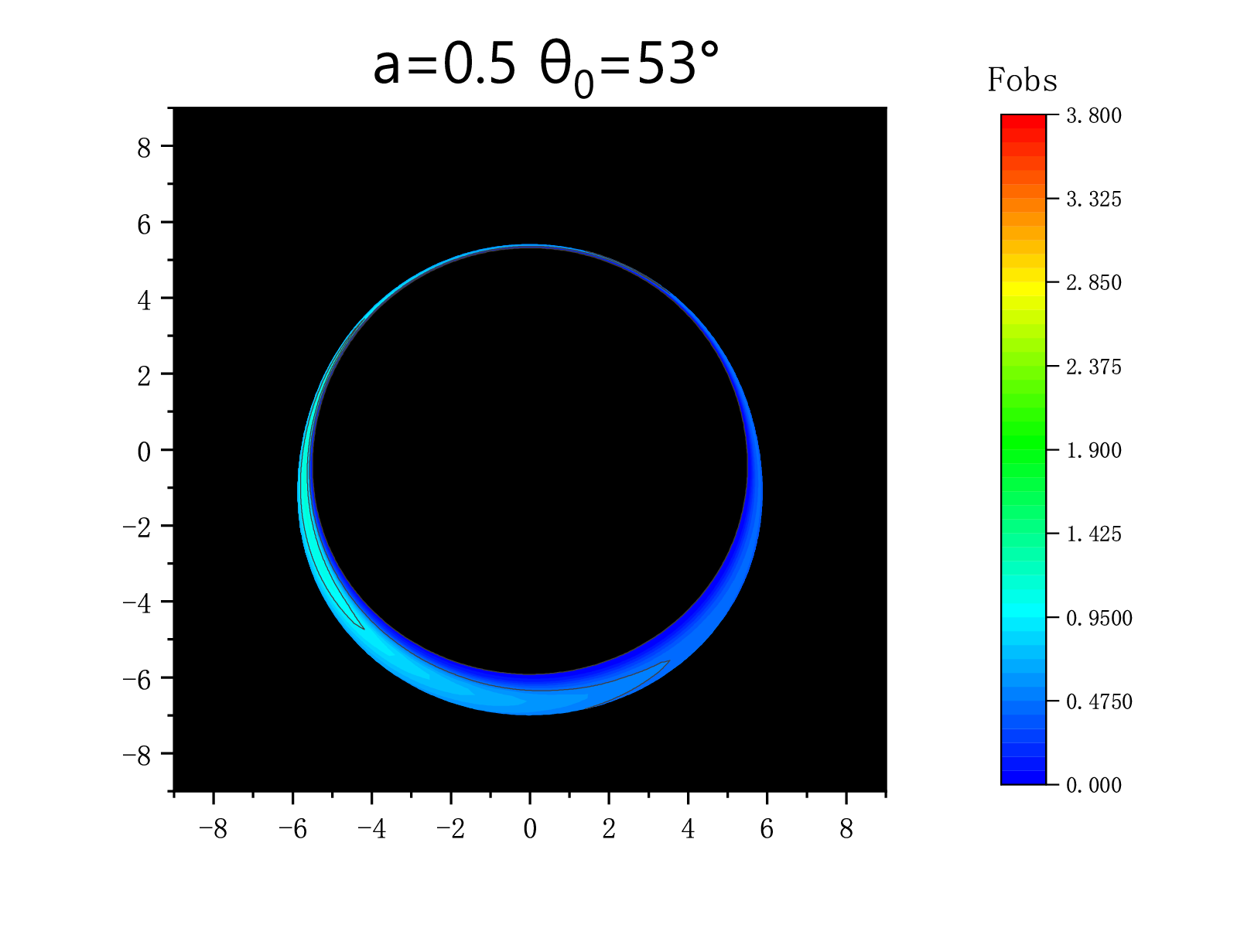}
  \hspace{0.5cm}
  \includegraphics[width=4.2cm,height=3.6cm]{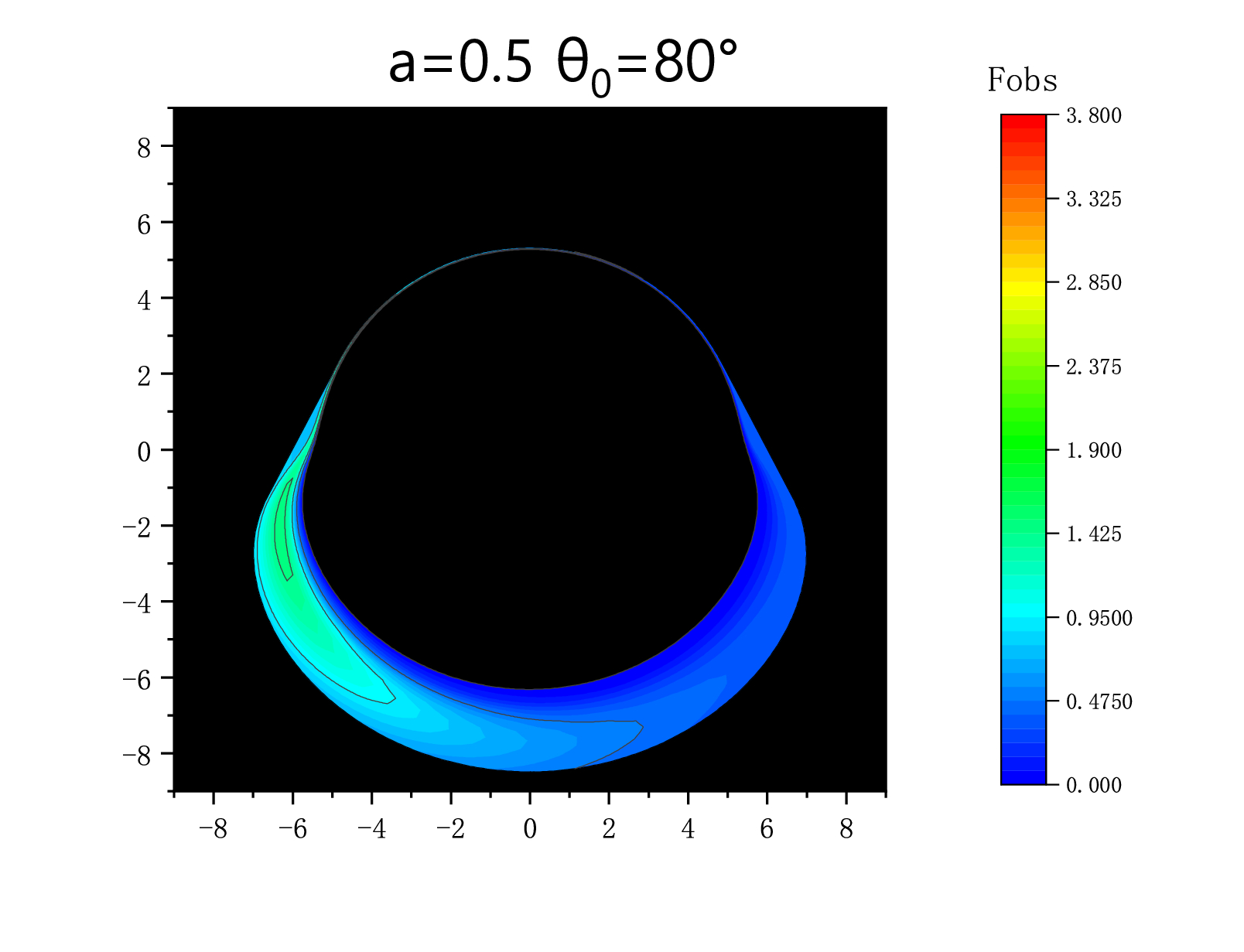}
\parbox[c]{15.0cm}{\footnotesize{\bf Fig~8.}  
The observed flux $F_{\rm obs}$ of the KS secondary image with the inclination angles of the observer $\theta_{0}=17^{\circ}$, $53^{\circ}$, and $80^{\circ}$. The inner edge of the disk at $r_{\rm in} = r_{\rm isco}$, and the outer edge of the disk is at $r= 15M$. {\em Top Panel}: the deformation parameter $a = 0.1$ and {\em Bottom Panel}: the deformation parameter $a = 0.5$. The BH mass is taken as $M = 1$.}
\label{fig8}
\end{center}

Figures 7 and 8 depict the observed flux for the direct and secondary images, respectively. It is evident that when the inclination angle of the observer ($\theta_{0}$) increases, the flux ($F_{\rm obs}$) becomes increasingly asymmetric. And at the same inclination angle, the increase of deformation parameter $a$ also leads to a decrease in flux. These phenomenons can be attributed to several factors. As $a$ increases, the gravitational field of the BH decreases, leading to changes in the velocity and density distribution of matter as it moves toward BH, which affects the amount of energy needed for light to escape from the BH, thus leading to the radiant flux distribution observed by distant observer. Secondly, due to the rotation of the accretion disk and the increase of the observation angle, the influence of the direction of light movement on the observer increases. Since the light on the left side of the disk moves towards the observer, while the right side moves away from the observer, the observed flux on the left side is larger. In summary, the observed radiant flux is affected by both gravitational redshift and Doppler effect. Additionally, an increase in the deformation parameter ($a$) results in a larger range of flux decreases.

\subsection{Data analysis}
\label{sec:4-2}
\par
To demonstrate the relationship between $F_{\rm obs}$ and $a$ more intuitively, we use the same method that produced Fig. 5. Fig. 9 illustrates that, under different observer angles $\theta_{0}$ ($\theta_{0}=17^{\circ}$, $53^{\circ}$, and $80^{\circ}$). Obviously, the accurate value (indicated by the red dotted line) reveals a gradual decrease in both the radiation flux of the observer $F_{\rm obs}$ and the intrinsic flux of the source $F$ as the deformation parameter $a$ increases in the range of 0 to 1. Furthermore, the rate of decrease accelerates progressively.

\begin{center}
 \includegraphics[width=4.2cm,height=3.6cm]{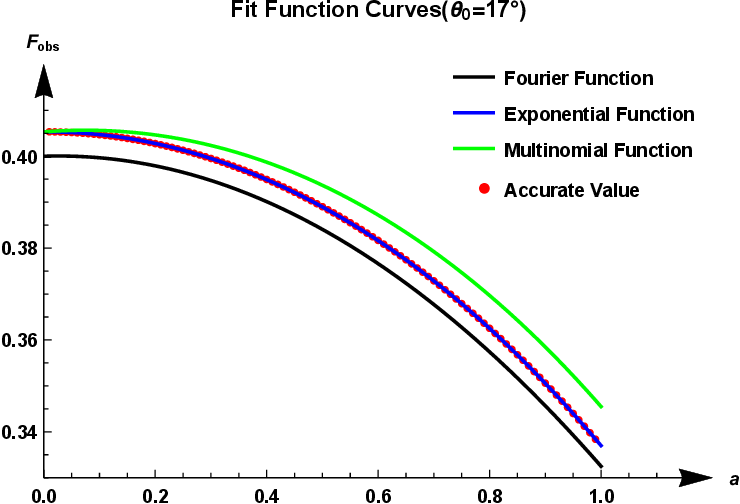}
  \hspace{0.5cm}
  \includegraphics[width=4.2cm,height=3.6cm]{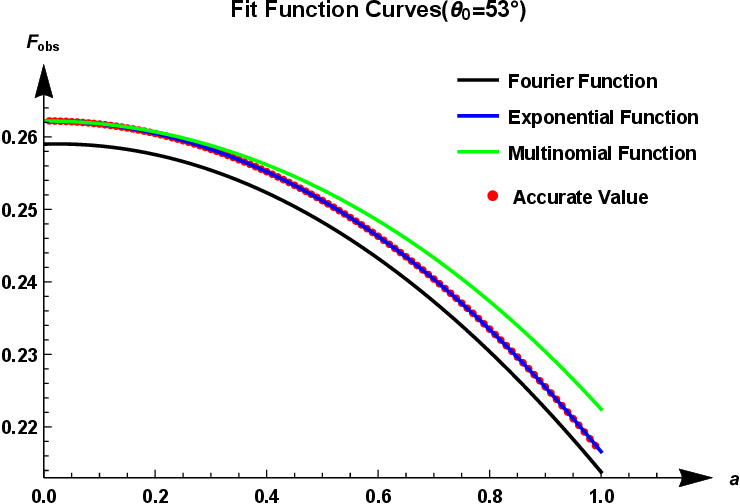}
  \hspace{0.5cm}
  \includegraphics[width=4.2cm,height=3.6cm]{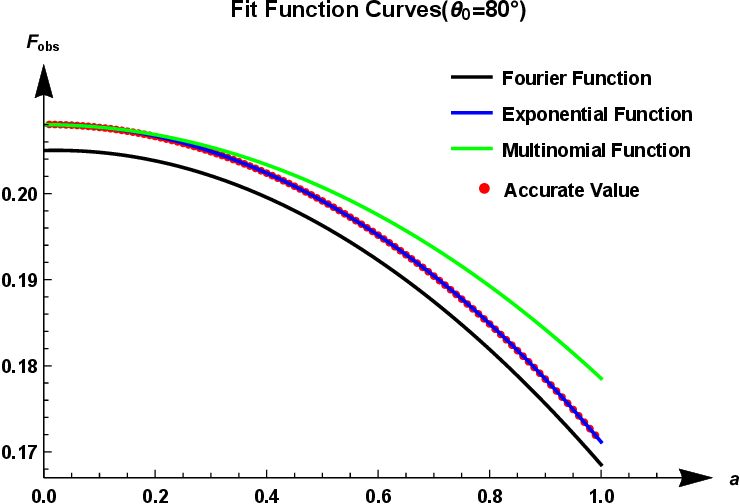}
\parbox[c]{15.0cm}{\footnotesize{\bf Fig~9.}  
Fitting of fourier function (the black solid line), exponential function(the blue solid line), and multinomial function (the green solid line) to the accurate value of the intrinsic flux of the observed flux $F_{\rm obs}$ and the deformation parameter $a$. {\em Left Panel}: The observer $\theta_{0}=17^{\circ}$. {\em Middle Panel}: The observer $\theta_{0}=53^{\circ}$. {\em Right Panel}: The observer $\theta_{0}=80^{\circ}$.}
\label{fig9}
\end{center}

\par
We assume that three functions are
\begin{eqnarray}
\label{4-7}
&&F=a_{1}e^{-(\frac{a-a_{2}}{a_{3}})^{2}}+a_{4}e^{-(\frac{a-a_{5}}{a_{6}})^{2}},\\
&&F=b_{1}+b_{2}\cos(a b_{3})+b_{4}\sin(a b_{5}),\\
&&F=c_{1}a^{6}+c_{2}a^{5}+c_{3}a^{4}+c_{4}a^{3}+c_{5}a^{2}+c_{6}a+c_{7},
\end{eqnarray}
where $a$, $b$, and $c$ with subscripts are the coefficient in the fitting function. Table 1 presents the numerical relationship between the intrinsic flux of the source ($F$), the observed flux ($F_{\rm obs}$), and the deformation parameter ($a$) at various angles ($\theta_{0}$). We fit these data with three different functions assumed above: a Fourier function (represented by the black solid line), an exponential function (represented by the blue solid line), and a multinomial function (represented by the green solid line).

\par
The Mean Squared Error (MSE) is a commonly used metric to assess the discrepancy between predicted and actual values in different prediction models. By calculating the squared differences between predicted and true values, the MSE provides a reliable measure of the accuracy of predictive models. In the present investigation, we have computed the MSE between the three functions and the true value, in order to determine the best fitting function. As shown in Table 2, the exponential function has an extremely small magnitude of MSE in all cases, indicating that it is the closest to the theoretical results. Conversely, the magnitudes of the Fourier and multinomial functions are relatively large, suggesting that their results deviate from the accurate value. Based on these results, we consider the exponential function as an empirical function for the radiation flux and the deformation parameter, and the parameter values for its fitting formula under various circumstances are presented in Table 3. Our analysis revealed that the exponential function can effectively describe the functional relationship between the observed flux and quantum deformation parameter, providing new insights into the influence of quantum correction on the observation results. Our results could serve as a valuable tool for understanding quantum gravity.
\begin{table*}
\caption{Numerical relationship between observed flux and the deformation parameter for $\alpha=\frac{\pi}{2}$ and $r$=15}
\label{Tab:1}
\begin{tabular}{c|cccc}
  $a$ &${F}$ &${F_{\rm obs}(\theta_{0}=17^{\circ})}$ &${F_{\rm obs}(\theta_{0}=53^{\circ})}$ &${F_{\rm obs}(\theta_{0}=80^{\circ})}$ \\
  \hline
  $0.05$ &$0.833439$   &$0.405174$   &$0.262100$   &$0.207945$   \\
  \hline
  $0.10$ &$0.832546$   &$0.404692$   &$0.261774$   &$0.207748$   \\
 \hline
  $0.15$ &$0.831055$   &$0.403887$   &$0.261231$   &$0.207242$   \\
  \hline
  $0.20$ &$0.828963$   &$0.402758$   &$0.260469$   &$0.206625$   \\
  \hline
  $0.25$ &$0.826263$   &$0.401303$   &$0.259702$   &$0.205831$   \\
  \hline
  $0.30$ &$0.822950$   &$0.399519$   &$0.258286$   &$0.204858$   \\
  \hline
  $0.35$ &$0.819014$   &$0.397403$   &$0.256860$   &$0.203705$   \\
  \hline
  $0.40$ &$0.814445$   &$0.394951$   &$0.255209$   &$0.202652$   \\
  \hline
  $0.45$ &$0.809232$   &$0.392158$   &$0.253330$   &$0.200852$   \\
  \hline
  $0.50$ &$0.803361$   &$0.389020$   &$0.251220$   &$0.199147$   \\
  \hline
  $0.55$ &$0.796816$   &$0.385529$   &$0.248876$   &$0.197254$   \\
  \hline
  $0.60$ &$0.789582$   &$0.381680$   &$0.246294$   &$0.19517$   \\
  \hline
  $0.65$ &$0.781638$   &$0.377464$   &$0.243469$   &$0.192891$   \\
  \hline
  $0.70$ &$0.772964$   &$0.372875$   &$0.240398$   &$0.194150$   \\
  \hline
  $0.75$ &$0.763537$   &$0.367903$   &$0.237075$   &$0.187737$   \\
  \hline
  $0.80$ &$0.753332$   &$0.362539$   &$0.233494$   &$0.184852$   \\
  \hline
  $0.85$ &$0.742322$   &$0.356772$   &$0.229650$   &$0.181760$   \\
  \hline
  $0.90$ &$0.730477$   &$0.350590$   &$0.225536$   &$0.178452$   \\
  \hline
  $0.95$ &$0.717766$   &$0.343983$   &$0.221145$   &$0.174924$   \\
\end{tabular}
\end{table*}
\begin{table*}
\caption{MSE of Fourier functions, exponential functions, and polynomial functions with accurate values for $\alpha=\frac{\pi}{2}$ and $r$=15}
\label{Tab:2}
\begin{tabular}{c|cccc}
  $-$ &${F}$ &${F_{\rm obs}(\theta_{0}=17^{\circ})}$ &${F_{\rm obs}(\theta_{0}=53^{\circ})}$ &${F_{\rm obs}(\theta_{0}=80^{\circ})}$ \\
  \hline
  $Fourier Function$ &$6.1196\times10^{-5}$     &$2.46304\times10^{-5}$   &$8.91666\times10^{-6}$    &$8.25137\times10^{-6}$   \\
  \hline
  $Exponential Function$ &$4.64208\times10^{-9}$       &$1.02525\times10^{-9}$       &$7.94073\times10^{-10}$     &$8.33254\times10^{-10}$   \\
  \hline
  $Multinomial Function$ &$9.21466\times10^{-5}$       &$2.67287\times10^{-5}$       &$7.06738\times10^{-6}$     &$9.61621\times10^{-6}$  \\
\end{tabular}
\end{table*}

\begin{table*}
\caption{Coefficient value of exponential fitting function under different conditions for $\alpha=\frac{\pi}{2}$ and $r$=15}
\label{Tab:3}
\begin{tabular}{c|cccccc}
  $-$ &${a_{1}}$ &${a_{2}}$ &${a_{3}}$ &${a_{4}}$ &${a_{5}}$ &${a_{6}}$\\
  \hline
  ${F}$ &$0.8336$     &$0.01004$   &$2.561$    &$-0.02833$  &$1.349$    &$0.4081$ \\
  \hline
  ${F_{\rm obs}(\theta_{0}=17^{\circ})}$ &$0.4053$      &$0.007409$       &$2.451$     &$-0.01798$  &$1.456$   &$0.4693$ \\
  \hline
  ${F_{\rm obs}(\theta_{0}=53^{\circ})}$  &$0.2622$  &$0.004475$  &$2.422$  &$-0.0102$  &$1.407$  &$0.4711$  \\
  \hline
  ${F_{\rm obs}(\theta_{0}=80^{\circ})}$ &$0.208$  &$0.01325$  &$2.341$  &$-0.007433$  &$1.383$  &$0.4006$  \\
\end{tabular}
\end{table*}

\begin{center}
 \includegraphics[width=4.2cm,height=3.6cm]{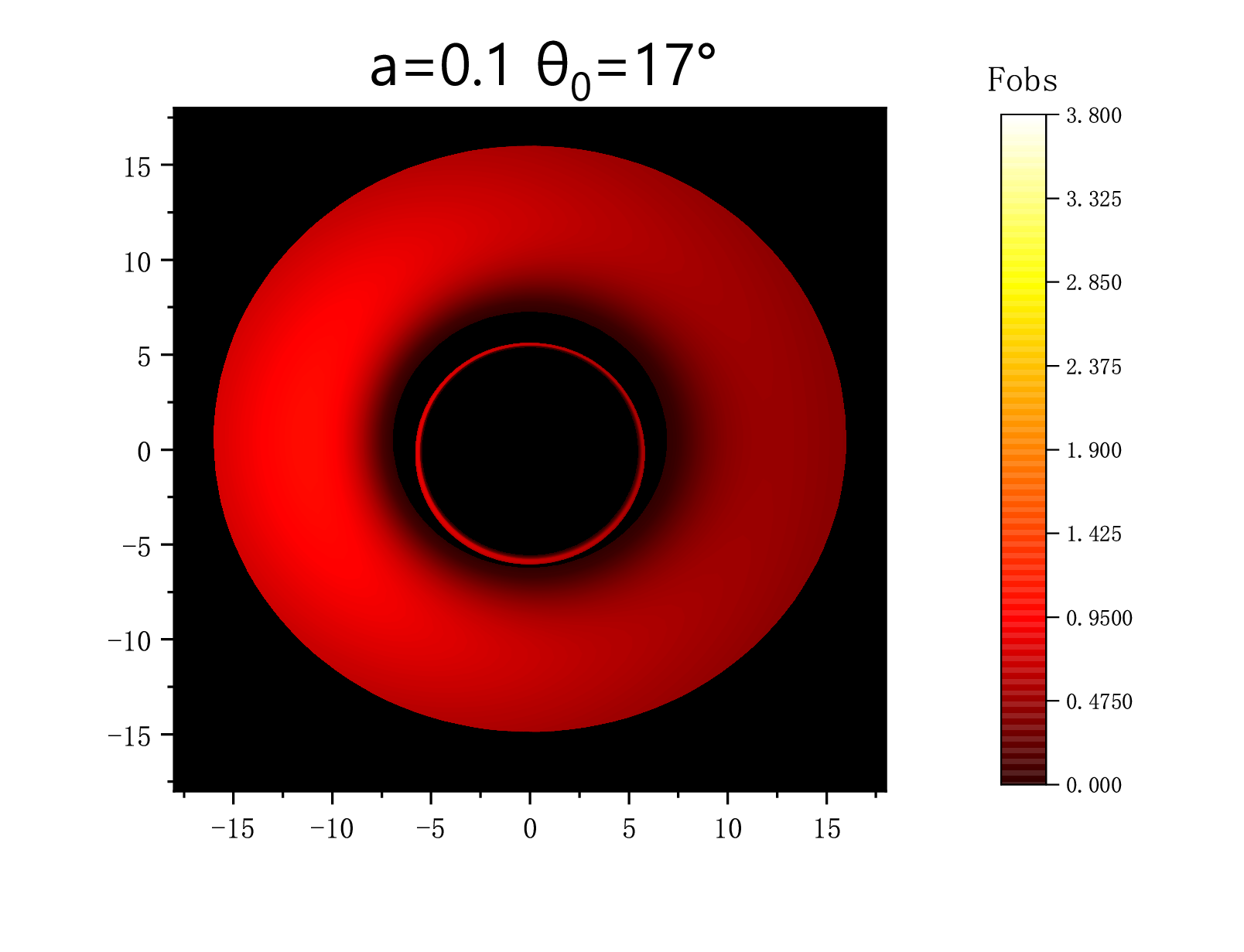}
  \hspace{0.5cm}
  \includegraphics[width=4.2cm,height=3.6cm]{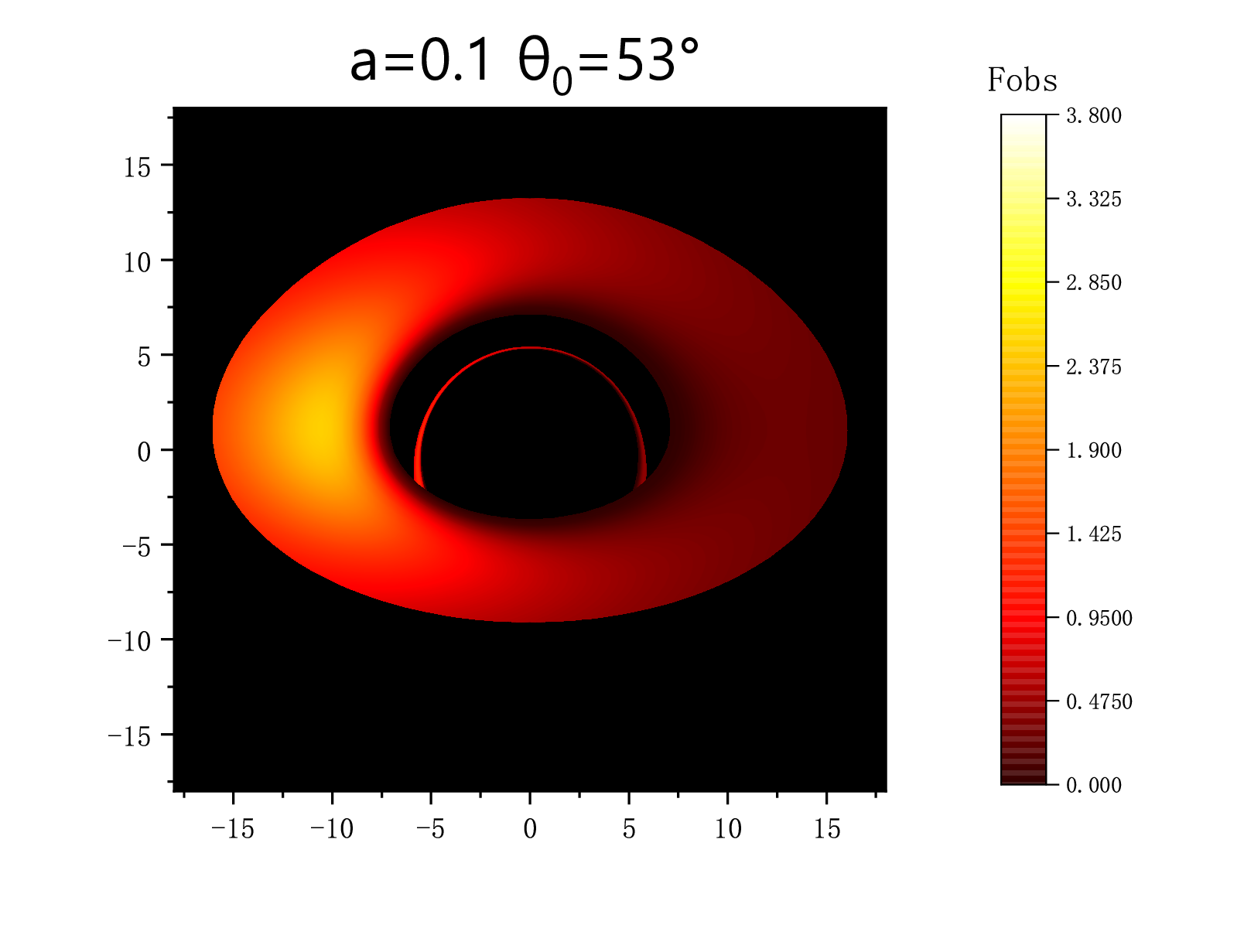}
  \hspace{0.5cm}
  \includegraphics[width=4.2cm,height=3.6cm]{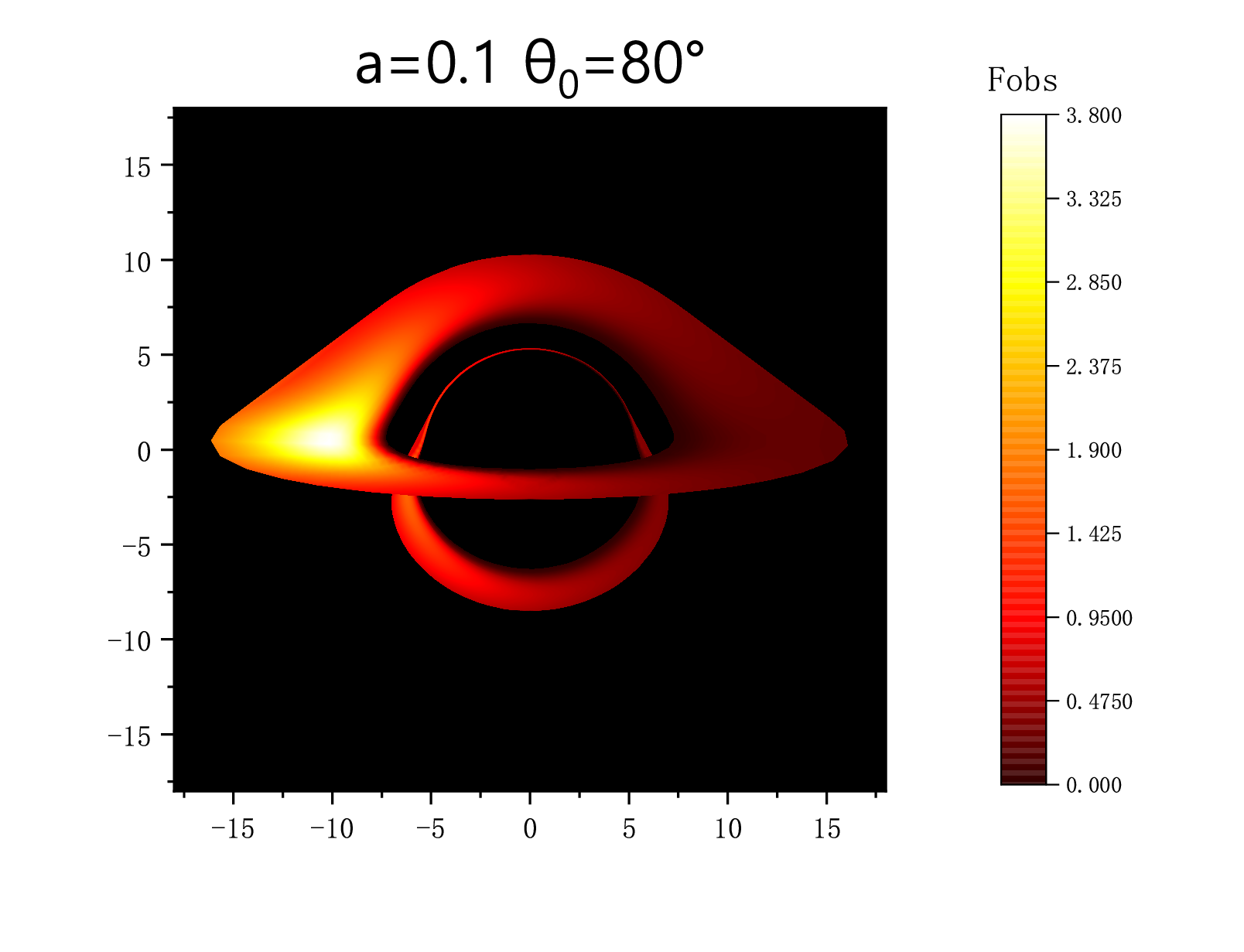}
  \hspace{0.5cm}
  \includegraphics[width=4.2cm,height=3.6cm]{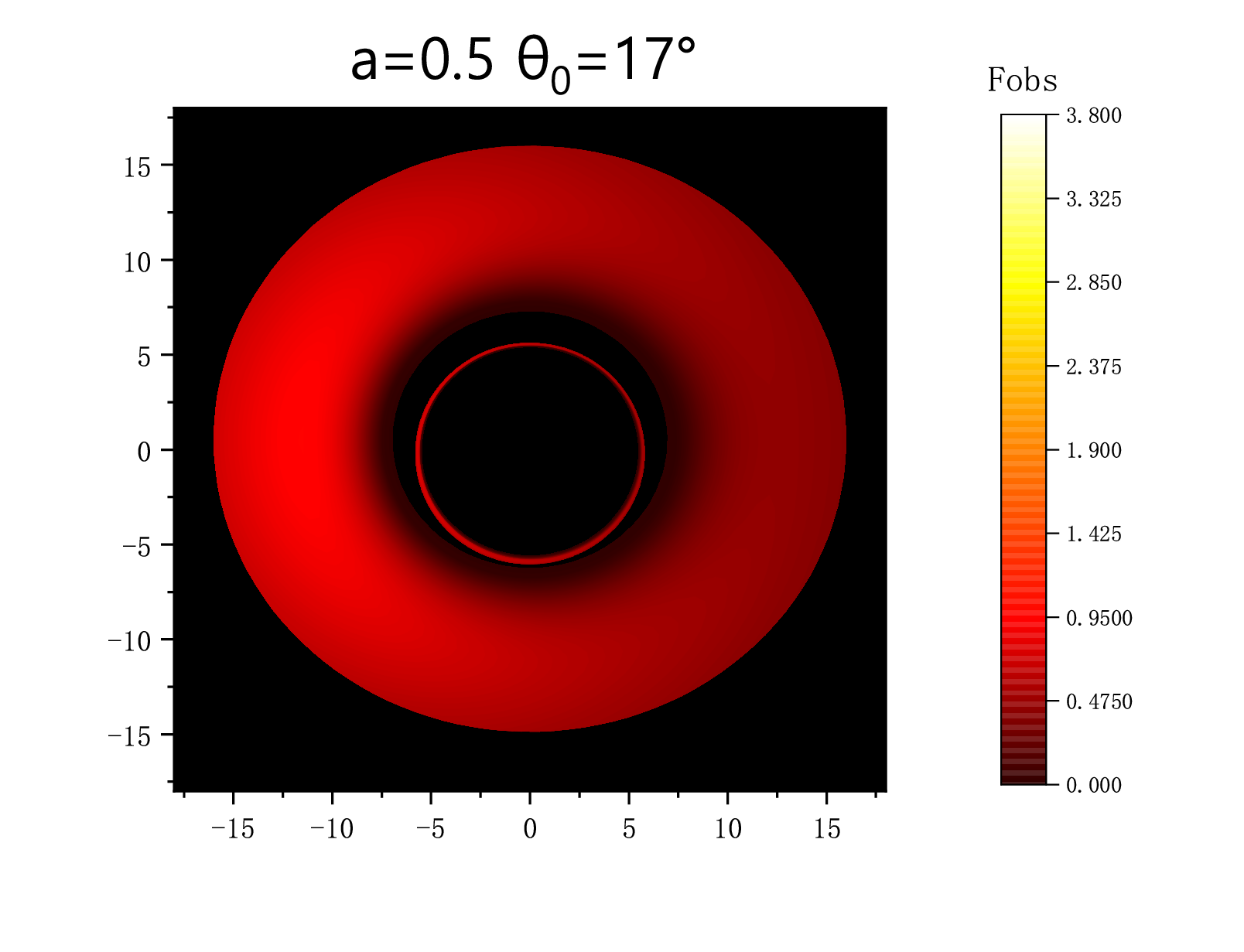}
  \hspace{0.5cm}
  \includegraphics[width=4.2cm,height=3.6cm]{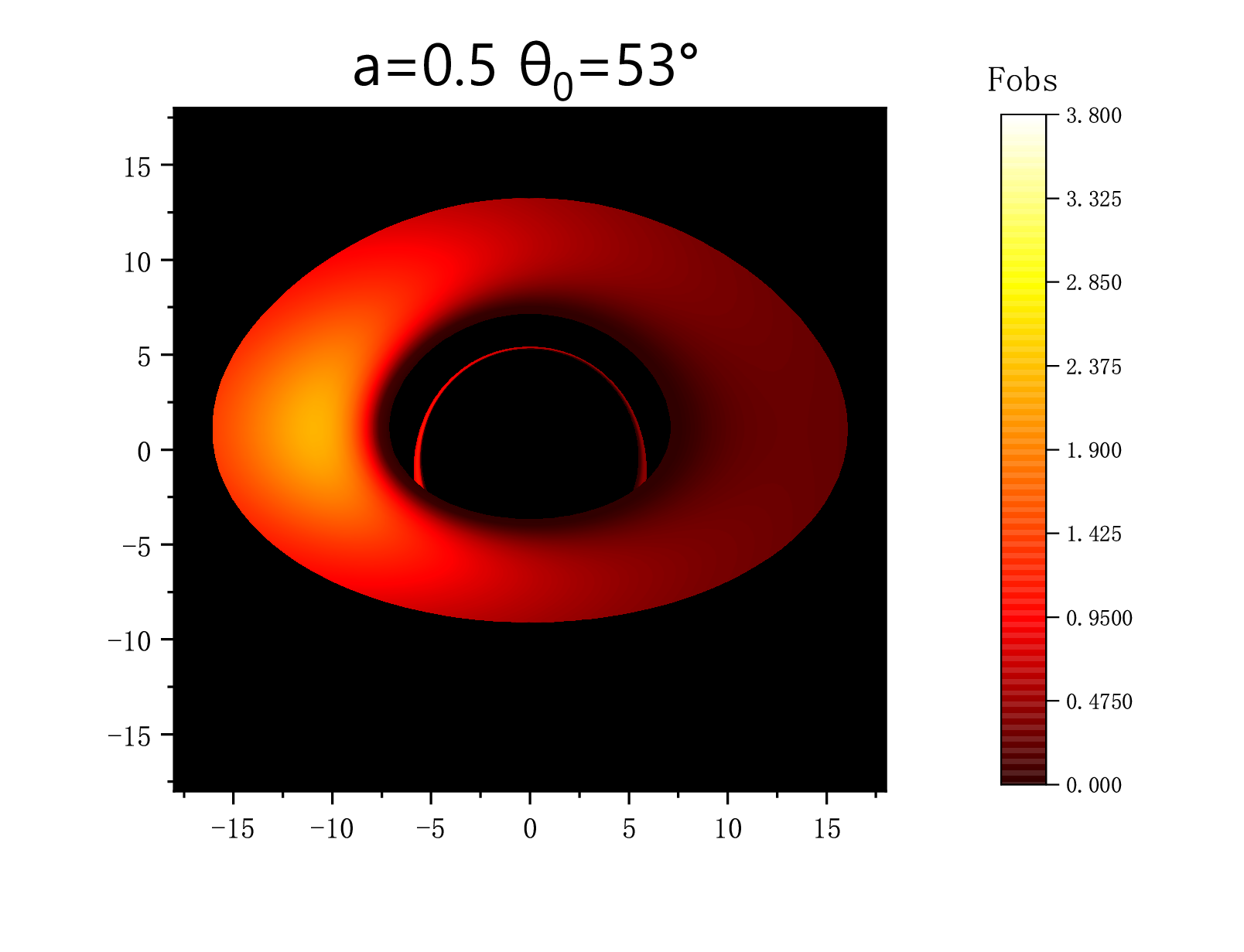}
  \hspace{0.5cm}
  \includegraphics[width=4.2cm,height=3.6cm]{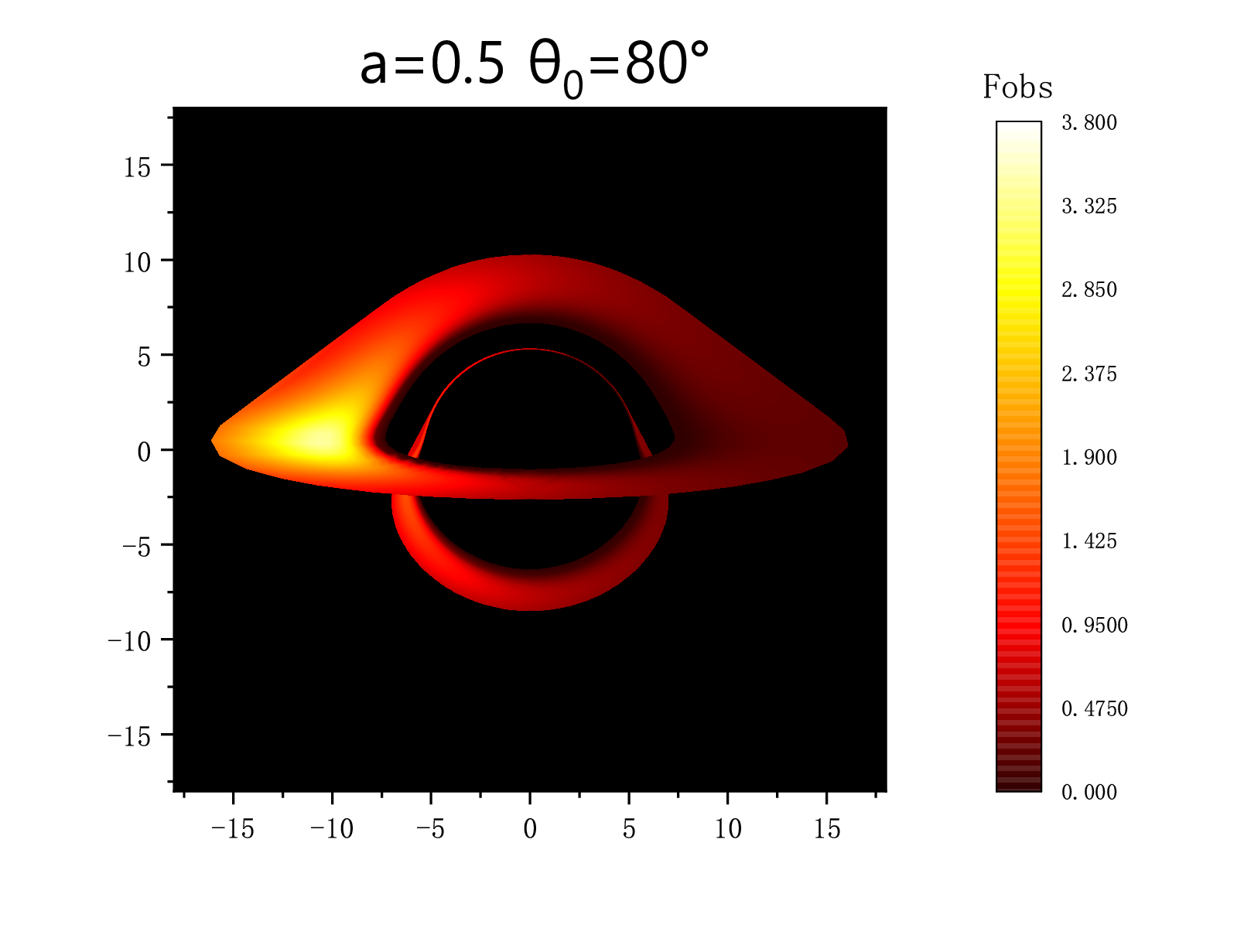}
\parbox[c]{15.0cm}{\footnotesize{\bf Fig~10.}  
Numerical simulation image of the KS BH is demonstrated at the inclination angles of the observer $\theta_{0}=17^{\circ}$, $53^{\circ}$, and $80^{\circ}$. The inner edge of the disk at $r_{\rm in} = r_{\rm isco}$, and the outer edge of the disk is at $r= 15M$. Top Panel: the deformation parameter $a = 0.1$. Bottom Panel: the deformation parameter $a = 0.5$. The BH mass is taken as $M = 1$.}
\label{fig10}
\end{center}

\par
It is widely recognized that theoretical calculations of BH images can provide valuable guidance for the observation of the Event Horizon Telescope (EHT), which utilizes Very Long Baseline Interferometry (VLBI) technology to achieve high-resolution imaging. Nonetheless, there exist some discrepancies between the observation results of the EHT and the theoretical calculations of BH images. Notwithstanding this, there is a close relationship between the two, which not only enhances our understanding of the nature and properties of BHs but also inspires further research in this field. In this study, we employed numerical simulations to generate an accretion disk image of the KS BH.  Figure. 10 illustrates the predicted observation images of the KS BH under various observation angles. It is worth noting that these predicted results may provide valuable guidance for achieving the scientific goals of the EHT in future research.

\section{Conclusions}
\label{sec:5}
\par
In this analysis, we investigate the optical radiation emitted by the Kazakov-Solodukhin black hole in the context of an optically thick and geometrically thin accretion disk. We derive the effective potential and deflection angle of the KS BH using the four-momentum of photons. Our findings suggest that the results obtained by Luminet's semi-analytical method and full numerical integration are in agreement. Additionally, we use EHT observations to constrain the quantum correction parameter $a$, and our results indicate that for M87$^{}$, $a$ is constrained to be $a \leq 2.51$ within $1\sigma$ and $a \leq 3.39$ within $2\sigma$, while for Sgr A$^{}$, $a$ is constrained to be $a \leq 0.55$ within $1\sigma$ and $a \leq 1.27$ within $2\sigma$.

\par
Using elliptic integrals, we simulate the orbits and flux of the direct and secondary images of the KS BH accretion disk under various observation angles. Additionally, we observe that the radius of the photon orbit decreases as the value of $a$ decreases. The significant impact of observation inclination on observation results is also consistent with Cunningham's research. For the radiation flux of the accretion disk, our results show that an increase in the deformation parameter $a$ leads to a notable reduction in the flux for large values. As the tilt angle increases, the separation between the direct and secondary images becomes more prominent, and the image takes on a more distinct ``hat-like'' shape. We assume three hypothetical functional models to fit the flux results, and it is observed that the exponential function is more consistent with the results under specific fixed parameters.

\par
The radiation emitted from the accretion disk surrounding a BH is influenced by both gravitational redshift and Doppler effect. As a result, the radiation spectrum observed by a remote observer may vary depending on their position relative to the disk. We derive the redshift images of the accretion disk under different parameter values. Our findings reveal that as the observation angle $\theta_{0}$ increases, the range of the redshift expands gradually. The change in the deformation parameter $a$ has a small impact on the gravitational redshift $z$, and as a result, the change in the redshift image in the figure is not significant. As the inclination angle of the observer ($\theta_{0}$) increases, the flux ($F_{\rm obs}$) exhibits increasingly asymmetric. Additionally, an increase in the deformation parameter ($a$) results in a larger range of flux decreases.

\par
Finally, by fitting a functional model to the observed flux at various observation inclinations, we find that the exponential function can effectively describe the relationship between the observed flux and the quantum deformation parameter. These findings provide new insights into the impact of quantum correction on observation outcomes and can serve as a valuable tool for understanding quantum gravity.

\section*{Acknowledgments}
This work is supported by the National Natural Science Foundation of China (Grant No. 11805166) and the Sichuan Youth Science and Technology Innovation Research Team (Grant No. 21CXTD0038).

\clearpage

\end{CJK}
\end{document}